\documentclass[10pt,a4paper,english,10pt, twocolumn, twoside]{IEEEtran}
\usepackage[T1]{fontenc}
\usepackage[latin9]{inputenc}
\usepackage{color}
\usepackage{babel}
\usepackage{float}
\usepackage{amsthm}
\usepackage{amsmath}
\usepackage{amssymb}
\usepackage{graphicx}
\usepackage{setspace}
\usepackage{esint}
\usepackage[unicode=true,pdfusetitle,
 bookmarks=true,bookmarksnumbered=true,bookmarksopen=false,
 breaklinks=false,pdfborder={0 0 1},backref=false,colorlinks=true]
 {hyperref}
\hypersetup{
 linkcolor=blue, citecolor=red}

\makeatletter

\floatstyle{ruled}
\newfloat{algorithm}{tbp}{loa}
\providecommand{\algorithmname}{Algorithm}
\floatname{algorithm}{\protect\algorithmname}

\theoremstyle{plain}
\newtheorem{thm}{\protect\theoremname}

\makeatother

\providecommand{\theoremname}{Theorem}

\begin{document}

\title{Leakage-Based Robust Beamforming for Multi-Antenna Broadcast System
with Per-Antenna Power Constraints and Quantized CDI
}

\begin{singlespace}

\author{\noindent {\normalsize Ming Ding$^{*}$, }\emph{\normalsize Member,
IEEE}{\normalsize , Meng Zhang, Hanwen Luo, and Wen Chen, }\emph{\normalsize Senior
Member, IEEE}%
\thanks{Copyright (c) 2012 IEEE. Personal use of this material is permitted.
However, permission to use this material for any other purposes must
be obtained from the IEEE by sending a request to \protect\href{http://pubs-permissions@ieee.org}{pubs-permissions@ieee.org}.%
}
\thanks{Ming Ding is with Sharp Laboratories of China Co., Ltd. (E-mail: ming.ding@cn.sharp-world.com).%
}
\thanks{Meng Zhang, Hanwen Luo and Wen Chen are with the Dept. of Electronic
Engineering, Shanghai Jiao Tong University, Shanghai, P. R. China.
(E-mail: \{mengzhang, hwluo, wenchen\}@sjtu.edu.cn). %
}
\thanks{This work is sponsored by Sharp Laboratories of China Co., Ltd.%
}}
\end{singlespace}

\maketitle
{}
\begin{abstract}
In this paper, we investigate the robust beamforming schemes for a
multi-user multiple-input-single-output (MU-MISO) system with per-antenna
power constraints and quantized channel direction information (CDI)
feedback. Our design objective is to maximize the expectation of the
weighted sum-rate performance by means of controlling the interference
leakage and properly allocating the power among user equipments (UEs).
First, we prove the optimality of the non-robust zero-forcing (ZF)
beamforming scheme in the sense of generating the minimum amount of
average inter-UE interference under quantized CDI. Then we derive
closed-form expressions of the cumulative density function (CDF) of
the interference leakage power for the non-robust ZF beamforming scheme,
based on which we adjust the leakage thresholds and propose two robust
beamforming schemes under per-antenna power constraints with an iterative
process to update the per-UE power allocations using the geometric
programming (GP). Simulation results show the superiority of the proposed
robust beamforming schemes compared with the existing schemes in terms
of the average weighted sum-rate performance. \end{abstract}

\section{Introduction}

Multi-antenna broadcast systems have gained considerable attention
as they can offer both spatial multiplexing and multi-user (MU) diversity
gains \cite{IntroBC}. In the multi-antenna broadcast channel (BC)
model, the multiplexing gain can be achieved by simultaneously serving
multiple user equipments (UEs) by space division multiple access schemes
with the dirty paper coding (DPC) \cite{DPC} or low-complexity linear
transmit beamforming, e.g., zero-forcing (ZF) beamforming \cite{ZF}.
Moreover, when the UE number is large, the capacity of the BC system
also grows with the UE number according to a double logarithm scaling
law due to the MU diversity gain \cite{multi-UE diversity gain}.
However, all these promising results are based on the assumption of
perfect channel direction information (CDI) available at the base
station (BS), which is too ideal for practical systems, especially
for the frequency division duplex (FDD) system such as the fourth
generation (4G) cellular network, e.g., the long term evolution advanced
(LTE-A) FDD system \cite{LTE-A}. The imperfectness of CDI is mainly
resulted from the limited-bit CDI quantization process performed by
the UE \cite{Limited-bit CSI}.

The existence of CDI quantization error motivates the design of robust
beamforming, which takes the uncertain channel distortions into account
{[}7-19{]}. In \cite{robust_minPower1}, the authors proposed robust
beamforming schemes for an MU multi-antenna BC system to minimize
the BS transmission power while maintaining certain quality-of-service
(QoS) requirements for the worst case model, i.e., treating channel
errors as norm bounded matrices, and the stochastic model, i.e., assuming
certain statistical properties of channel errors. The authors of \cite{basic_robust_SLNR}
addressed a more general problem by considering additional constraints,
such as keeping the interference under a preset tolerable level and
individually shaping the beamforming vectors. In \cite{robust_MMSE0},
the authors investigated robust beamforming schemes to minimize the
sum of UEs' mean squared errors (MSEs). Considering inter-UE interference
leakage \cite{origSINR}, the authors of \cite{robust_AveSLNR} designed
a robust beamforming scheme to maximize a lower bound of each UE's
average signal-to-leakage-plus-noise ratio (SLNR). Recently, in \cite{robust_ProbLeak1}
and \cite{robust_ProbLeak_ICC12}, the authors proposed another leakage-based
robust transmit beamforming scheme, which optimizes the average signal-to-interference-plus-noise
ratio (SINR) performance implicitly by maximizing the average signal
power subject to probabilistic leakage and noise power constraints.
Besides, in \cite{robust_interfchannel}, by taking finite-rate CDI
feedback into account the authors investigated the transceiver design
for a two-UE multiple-input multiple-output (MIMO) interference channel,
where each precoder or equalizer is divided into outer and inner parts
to eliminate the cross-link interference. In \cite{robust_MSE_new},
a joint design of the channel estimator and the quantization function
was proposed based on the criterion of MSE minimization. Furthermore,
robust beamforming schemes have been extended to more sophisticated
models such as the MIMO relay networks \cite{robust_extRelay1}, \cite{Prof_chen_relatWork1},
\cite{Prof_chen_relatWork2} and the multi-cell coordinated beamforming
operations \cite{robust_maxSINR1}, etc.

In this paper, we further investigate the robust beamforming design
based on the approach of leakage control. In particular, we consider
the optimization of the weighted sum-rate performance and more realistic
power constraints, which limit the BS transmission power on a per-antenna
basis. Compared with the existing robust beamforming schemes, our
assumption on the transmission power is more practical since each
antenna of a multi-antenna BS \cite{Per_ante_PC} is normally equipped
with an individual power amplifier at its analog front-end. Our contributions
are two folds:
\begin{enumerate}
\item \noindent We prove the optimality of the non-robust ZF beamforming
scheme in the sense of generating the minimum amount of average inter-UE
interference under limited-bit CDI, and derive closed-form expressions
of the cumulative density function (CDF) of the interference leakage
power for the non-robust ZF beamforming scheme.
\item \noindent We adjust the leakage thresholds based on the derived CDF
of the leakage power and propose a minimum average leakage control
(MALC) and a relaxed average leakage control (RALC) beamforming schemes
under per-antenna power constraints, together with an iterative process
to update the power allocation among UEs using the geometric programming
(GP) to maximize the weighted sum-rate performance.
\end{enumerate}
$\quad$Although in this paper we mainly treat the MU multiple-input-single-output
(MISO) model, our results can be extended to other BC models, such
as MIMO relay networks \cite{Future_ext_MIMOrelay2} and multi-cell
joint transmissions \cite{Future_ext_JT}.

\textit{Notations}: $\left(\cdot\right)^{\textrm{T}}$, $\left(\cdot\right)^{\textrm{H}}$,
$\left(\cdot\right)^{-1}$, $\left(\cdot\right)^{\dagger}$, $\textrm{tr}\left\{ \mathbf{\cdot}\right\} $
and $\textrm{rank}\left\{ \mathbf{\cdot}\right\} $ stand for the
transpose, conjugate transpose, inverse, pseudo-inverse, trace and
rank of a matrix, respectively. $\mathbf{I}_{N}$ stands for an $N\times N$
identity matrix. $\mathbf{A}_{i,:}$, $\mathbf{A}_{:,j}$ and $\mathbf{A}_{i,j}$
respectively denote the $i$-th row, $j$-th column and $\left(i,j\right)$-th
entry of matrix $\mathbf{A}$. Besides, $\mathbf{A}\succeq0$ and
$\mathbf{A}\in\mathbb{\mathbb{H}}_{N}^{\mathit{+}}$ mean that matrix
$\mathbf{A}$ is positive semi-definite and $\mathbf{A}$ is an $N$
by $N$ positive semi-definite Hermitian matrix, respectively. $\left\Vert \mathbf{a}\right\Vert $
and $\mathbf{a}_{i}$ denotes the Euclidean norm and $i$-th element
of vector $\mathbf{a}$. $\mathbb{E}_{\left[\mathbf{x}\right]}\left\{ \cdot\right\} $
and $\textrm{Re}\left\{ \cdot\right\} $ denote the expectation operation
over a random vector $\mathbf{x}$ and the real part of a complex
value. $\textrm{C}_{j}^{i}$ counts the combinations of choosing $i$
elements from a set of $j$ elements. $\mathcal{N}\left(\mathbf{0},\mathbf{X}\right)$
represents a circularly symmetric complex Gaussian distribution with
mean of zero vector and covariance matrix $\mathbf{X}$. Finally,
we denote $\Pr\left(x\right)$ as the probability of event $x$.

\section{System Model}

\begin{figure}[H]
\centering \includegraphics[width=5cm]{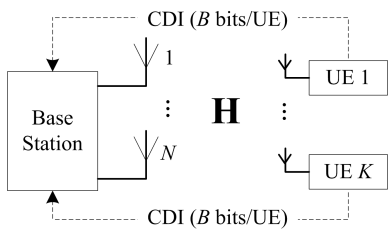} 
\\
\vspace{-0.5em}
\renewcommand{\figurename}{Fig.}

\caption{\label{fig:Sys_model}Illustration of a downlink MU-MISO system with
limited-bit CDI.}
\end{figure}

We consider a downlink MU-MISO system with limited-bit CDI feedback
as shown in Fig.~\ref{fig:Sys_model}, where a BS is equipped with
$N$ transmit antennas, and $K$ single-antenna UEs are served simultaneously.
To support $K$ independent data streams, it requires $N\geq K$.
However, our results can be easily extended to the case of $N<K$
with UE selection performed at the BS \cite{Prof_Chen_UE_selection}.

In Fig.~\ref{fig:Sys_model}, the large-scale channel attenuation
coefficient and the small-scale base-band channel vector between the
BS and the $k$-th $\left(k\in\left\{ 1,2,\ldots,K\right\} \right)$
UE are denoted as $\xi_{k}$ and $\mathbf{h}_{k}\in\mathbb{C}^{1\times N}$,
respectively. Let $\mathbf{w}_{j}\in\mathbb{C}^{\mathit{N}\times1}$
be the beamforming vector for UE $j$. Then the signal received at
UE $k$ can be described by

\begin{singlespace}
\noindent
\begin{equation}
y_{k}=\xi_{k}\mathbf{h}_{k}\sum_{j=1}^{K}\mathbf{w}_{j}x_{j}+n_{k}=\xi_{k}\mathbf{h}_{k}\mathbf{\mathbf{W}}\mathbf{\mathbf{x}}+n_{k},\label{eq:yk}
\end{equation}

\end{singlespace}

\noindent where $\mathbf{x}=\left[x_{1},\ldots,x_{k},\ldots,x_{K}\right]^{\textrm{T}}$
and $x_{k}$ is the data symbol intended for UE $k$. Without loss
of generality, we assume that $\mathbf{x}$ satisfies $\mathbb{E}_{\left[\mathbf{x}\right]}\left\{ \mathbf{x\mathbf{x^{\textrm{H}}}}\right\} =\mathbf{I}_{K}$.
The channel coefficients in $\mathbf{h}_{k}$ are assumed to experience
independently identical distribution (i.i.d.) Rayleigh flat fading
and remain unchanged during the downlink MU-MISO transmission. $n_{k}$
is a zero-mean circularly symmetric complex Gaussian (ZMCSCG) noise
variable with $\mathbb{E}_{\left[n_{k}\right]}\left\{ n_{k}n_{k}^{\textrm{H}}\right\} =N_{0}$.
In addition, $\mathbf{\mathbf{W}}=\left[\mathbf{w}_{1},\mathbf{w}_{2},\ldots,\mathbf{w}_{K}\right]$
and it is subject to an average per-antenna transmit power constraint
expressed as

\begin{singlespace}
\noindent
\begin{equation}
\mathbb{E}_{\left[\mathbf{x}\right]}\left\{ \left[\mathbf{\mathbf{W}}\mathbf{\mathbf{x}}\left(\mathbf{\mathbf{W}}\mathbf{\mathbf{x}}\right)^{\textrm{H}}\right]_{n,n}\right\} =\left[\mathbf{\mathbf{W}}\mathbf{\mathbf{W}}^{\textrm{H}}\right]_{n,n}\leq P_{n},\label{eq:PAPC}
\end{equation}

\end{singlespace}

\noindent where $P_{n}$ is the maximum transmission power of the
$n$-th $\left(n\in\left\{ 1,2,\ldots,N\right\} \right)$ antenna.
In addition, we denote the BS's maximum transmission power as $P=\sum_{n=1}^{N}P_{n}$.
By stacking the received signals of all UEs, we have

\begin{onehalfspace}
\noindent
\begin{equation}
\mathbf{y}=\mathbf{H}\mathbf{\mathbf{W}}\mathbf{\mathbf{x}}+\mathbf{n},\label{eq:vec_y}
\end{equation}

\end{onehalfspace}

\noindent where $\mathbf{y}=\left[y_{1},y_{2},\ldots,y_{K}\right]^{\textrm{T}}$,
$\mathbf{H}=\left[\xi_{1}\mathbf{h}_{1}^{\textrm{T}},\xi_{2}\mathbf{h}_{2}^{\textrm{T}},\ldots,\xi_{K}\mathbf{h}_{K}^{\textrm{T}}\right]^{\textrm{T}}$
and $\mathbf{n}=\left[n_{1},n_{2},\ldots,n_{K}\right]^{\textrm{T}}$.

The information of $\mathbf{H}$, i.e., the channel state information
(CSI), is composed of two parts, which are the channel direction information
(CDI) and channel magnitude information (CMI). The CDI is the normalized
base-band channel vector of UE $k$ denoted as $\mathbf{\tilde{h}}_{k}=\frac{\mathbf{h}_{k}}{\left\Vert \mathbf{h}_{k}\right\Vert }$
and the CMI is expressed by $\xi_{k}\left\Vert \mathbf{h}_{k}\right\Vert $
or $\xi_{k}^{2}\left\Vert \mathbf{h}_{k}\right\Vert ^{2}$. In practice,
perfect CSI is usually not available at the BS side. Hence, we first
assume imperfect CDI for the interested MU-MISO system shown in Fig.~\ref{fig:Sys_model},
where each UE quantizes its CDI and feeds it back to the BS with $B$
bits. The quantized CDI is defined as the index of a vector $\mathbf{\hat{h}}_{k}$
chosen from a random vector quantization (RVQ) codebook $\mathbf{C}_{k}=\left\{ \mathbf{c}_{k,1},\mathrm{\mathbf{c}}_{k,2},\ldots,\mathrm{\mathbf{c}}_{k,2^{B}}\right\} $
\cite{RVQ} to match $\mathbf{\tilde{h}}_{k}$. To be more specific,
$\mathbf{C}_{k}$ consists of $2^{B}$ unit vectors $\mathbf{c}_{k,n}$
$\left(n\in\left\{ 1,2,\ldots,2^{B}\right\} \right)$ isotropically
distributed in $\mathbb{C}^{1\times M}$ and $\mathbf{\hat{h}}_{k}$
is selected as

\begin{equation}
\mathbf{\hat{h}}_{k}=\underset{\mathbf{c}_{k,n}\in\mathbf{C}_{k}}{\arg\max}\left|\mathbf{c}_{k,n}\mathbf{\tilde{h}}_{k}^{\textrm{H}}\right|.\label{eq:hk_hat}
\end{equation}

\noindent \begin{flushleft}
Then $\mathbf{\tilde{h}}_{k}$ can be decomposed as
\par\end{flushleft}

\begin{singlespace}
\noindent
\begin{equation}
\mathbf{\tilde{h}}_{k}=\cos\left(\angle\left(\mathbf{\tilde{h}}_{k},\mathbf{\hat{h}}_{k}\right)\right)\mathbf{\hat{h}}_{k}+\sin\left(\angle\left(\mathbf{\tilde{h}}_{k},\mathbf{\hat{h}}_{k}\right)\right)\mathbf{e}_{k},\label{eq:decomp_hk_tilde}
\end{equation}

\end{singlespace}

\noindent where $\mathbf{e}_{k}$ is a quantization error vector orthogonal
to $\mathbf{\hat{h}}_{k}$.

Regarding the CMI, $\xi_{k}$ or $\xi_{k}^{2}$ can be inferred from
UE $k$'s pathloss information, i.e., $\frac{1}{\xi_{k}^{2}}$, which
is implicitly reported to the BS for mobility management in existing
cellular networks \cite{On_LTE}. In addition, $\left\Vert \mathbf{h}_{k}\right\Vert $
or $\left\Vert \mathbf{h}_{k}\right\Vert ^{2}$can be easily quantized
using $M$ bits as long as its CDF is known. Since we consider the
Rayleigh fading channels in this paper, $\left\Vert \mathbf{h}_{k}\right\Vert ^{2}$
follows a chi-squared distribution with its CDF expressed as $P_{R}\left(r\right)=1-\textrm{e}^{-r}\sum_{l=0}^{N-1}\frac{r^{l}}{l!}$
\cite{Proakis book}. Thus we can divide $P_{R}\left(r\right)$ into
$2^{M}$ segments and select the midpoint of each segment to construct
the codebook for the quantization of $\left\Vert \mathbf{h}_{k}\right\Vert ^{2}$
as $\left\{ \left.T_{i}=P_{R}^{-1}\left(\frac{2i+1}{2^{M+1}}\right)\right|i\in\left\{ 0,1,\ldots,2^{M}-1\right\} \right\} $.
Then the quantized $\left\Vert \mathbf{h}_{k}\right\Vert ^{2}$ is
given by

\noindent
\begin{equation}
\hat{A}_{k}=\underset{T_{i}}{\arg\min}\left(\left|\left\Vert \mathbf{h}_{k}\right\Vert ^{2}-T_{i}\right|\right).\label{eq:hk_CF_CMI}
\end{equation}
In the sequel, $\xi_{k}^{2}$ will be referred to as the pathloss
CMI (PL-CMI) and $\left\Vert \mathbf{h}_{k}\right\Vert ^{2}$ as the
channel fading CMI (CF-CMI). The PL-CMI is assumed to be perfect because
it changes very slowly, and hence its quantization accuracy can be
consistently improved over a long period of time. The CF-CMI, on the
other hand, should be subject to quantization errors since it varies
as fast as the CDI. Fortunately, in our simulations, we will show
that even with the average CF-CMI only, i.e., $\mathbb{E}_{\left[\mathbf{h}_{k}\right]}\left\{ \left\Vert \mathbf{h}_{k}\right\Vert ^{2}\right\} $,
the performance of the interested beamforming schemes is comparable
with that achieved by the perfect CF-CMI, i.e., $\left\Vert \mathbf{h}_{k}\right\Vert ^{2}$.
For notational brevity, we denote $A_{k}^{\textrm{ave}}=\mathbb{E}_{\left[\mathbf{h}_{k}\right]}\left\{ \left\Vert \mathbf{h}_{k}\right\Vert ^{2}\right\} $
hereafter. Note that $A_{k}^{\textrm{ave}}$ can be obtained through
analytical calculation or numerical simulation based on the assumption
of the channel model. Thus, $A_{k}^{\textrm{ave}}$ is not required
to be fed back to the BS, i.e., $M=0$ for the average CF-CMI case.
In the following, we will concentrate on the quantized CDI $\mathbf{\hat{h}}_{k}$
with $B>0$ and assume average CF-CMI at the BS. We will explicitly
state if quantized CF-CMI $\hat{A}_{k}$ is considered.

\section{Non-Robust Beamforming Schemes}

For non-robust beamforming schemes, the channel uncertainties due
to the CDI quantization errors are ignored so that at the BS side
each UE's channel vector is directly replaced by $\mathbf{\check{h}}_{k}=\sqrt{A_{k}^{\textrm{ave}}}\mathbf{\hat{h}}_{k}$.
First, we discuss the non-robust ZF beamforming scheme with per-antenna
power constraints, which results will serve as the benchmark for performance
comparison in our simulations.

\subsection{The Non-Robust ZF Beamforming Scheme with Per-antenna Power Constraints}

A commonly used beamforming scheme for the downlink MU-MISO system
is based on the ZF approach, which aims to completely mitigate the
inter-UE interference. Under the sum power constraint, the non-robust
ZF beamforming vector for UE $k$ can be derived as \cite{ZF}

\begin{singlespace}
\noindent
\begin{equation}
\mathbf{w}_{k}^{\textrm{ZF}}=\sqrt{\tilde{P}_{k}}\mathbf{\tilde{w}}_{k}^{\textrm{ZF}}=\sqrt{\tilde{P}_{k}}\frac{\mathbf{\check{H}}_{:,k}^{\dagger}}{\left\Vert \mathbf{\check{H}}_{:,k}^{\dagger}\right\Vert },\label{eq:NR_ZF_wk}
\end{equation}

\end{singlespace}

\noindent where $\tilde{P}_{k}$ is the transmission power for UE
$k$ and it satisfies the sum power constraint $\sum_{k=1}^{K}\tilde{P}_{k}\leq P$.
Besides, $\mathbf{\tilde{w}}_{k}^{\textrm{ZF}}$ is the normalized
precoding vector and $\mathbf{\check{H}}^{\dagger}$ is the pseudo-inverse
of $\mathbf{\check{H}}=\left[\xi_{1}\mathbf{\check{h}}_{1}^{\textrm{T}},\xi_{2}\mathbf{\check{h}}_{2}^{\textrm{T}},\ldots,\xi_{K}\mathbf{\check{h}}_{K}^{\textrm{T}}\right]^{\textrm{T}}$.
In the following, the non-robust ZF beamforming scheme based on \eqref{eq:NR_ZF_wk}
will be referred to as the non-robust ZF scheme or the ZF scheme for
short.

It is obvious that $\mathbf{w}_{k}^{\textrm{ZF}}$ may violate the
per-antenna power constraints shown in \eqref{eq:PAPC}. In \cite{ZF_per_ante_PC},
the authors investigated the ZF beamforming with per-antenna power
constraints and demonstrated that the optimal solution depends on
the specific objective function. Considering the maximization of the
weighted sum-rate for a MISO system, the authors of \cite{ZF_per_ante_PC}
addressed that the non-robust beamforming solution can be found by
solving a standard semi-definite program (SDP) problem shown as

\begin{singlespace}
\noindent
\begin{eqnarray}
\underset{\mathbf{Q}_{k}\in\mathbb{\mathbb{H}}_{N}^{\mathit{+}}}{\max} &  & f\left(\left\{ \mathbf{Q}_{k}\right\} \right)=\sum_{k=1}^{K}\alpha_{k}\log_{2}\left(1+\frac{\xi_{k}^{2}\mathbf{\check{h}}_{k}\mathbf{Q}_{k}\mathbf{\check{h}}_{k}^{\textrm{H}}}{N_{0}}\right)\nonumber \\
\textrm{s.t.} &  & \textrm{tr}\left\{ \mathbf{Q}_{k}\mathbf{\mathbf{\check{h}}}_{j}^{\textrm{H}}\mathbf{\mathbf{\check{h}}}_{j}\right\} =0,\quad\forall j\neq k;\nonumber \\
 &  & \sum_{k=1}^{K}\left[\mathbf{Q}_{k}\right]_{n,n}\leq P_{n},\quad\forall n,\label{eq:Problem_ZF_PA}
\end{eqnarray}

\end{singlespace}

\noindent where $\mathbf{Q}_{k}\in\mathbb{\mathbb{H}}_{N}^{\mathit{+}}$
denotes that $\mathbf{Q}_{k}$ is an $N$ by $N$ positive semi-definite
Hermitian matrix. The first and second sets of constraints in problem
\eqref{eq:Problem_ZF_PA} represent the requirements of zero interference
among UEs and per-antenna power limitations, respectively. Problem
\eqref{eq:Problem_ZF_PA} is a convex optimization problem and its
numerical solution can be obtained by the use of standard mathematical
softwares \cite{cvx_software}. Note that it has been proven in \cite{ZF_per_ante_PC}
that problem \eqref{eq:Problem_ZF_PA} always admits a solution with
rank-one matrices. Thus, the rank-one constraints on $\left\{ \mathbf{Q}_{k}\right\} $
for beamforming operations have been omitted. Suppose that $\left\{ \mathbf{Q}_{k}^{\textrm{ZF-PA}}\right\} $
is the solution to problem \eqref{eq:Problem_ZF_PA} and $\mathbf{Q}_{k}^{\textrm{ZF-PA}}=\mathbf{q}_{k}^{\textrm{ZF-PA}}\left(\mathbf{q}_{k}^{\textrm{ZF-PA}}\right)^{\textrm{H}}$,
then the non-robust ZF beamforming vector for UE $k$ under per-antenna
power constraints becomes

\begin{singlespace}
\noindent
\begin{equation}
\mathbf{w}_{k}^{\textrm{ZF-PA}}=\mathbf{q}_{k}^{\textrm{ZF-PA}}.\label{eq:wk_ZF-PA}
\end{equation}

\end{singlespace}

\noindent The non-robust beamforming scheme under per-antenna power
constraints with the solution of \eqref{eq:wk_ZF-PA} will be called
as the ZF-PA scheme for abbreviation. It should be noted that both
the ZF and ZF-PA schemes will suffer from large performance degradation
due to unknown inter-UE interference resulted from inevitable CDI
quantization errors \cite{robust_AveSLNR}, \cite{robust_ProbLeak_ICC12}.

\subsection{The Non-Robust SLNR Beamforming Scheme with Per-UE Power Constraints}

The ZF approach merely focuses on the minimization of the inter-UE
interference power, whereas the signal power usually suffers from
considerable power loss due to forced vector steering away from the
subspace spanned by other UEs' channel vectors. Therefore, an alternative
SLNR approach has been proposed in \cite{origSINR}, which suggests
maximizing the ratio of the signal power over the interference leakage
power plus the noise power. The non-robust SLNR beamforming scheme
under per-UE power constraints, referred to as the SLNR scheme for
short, generates the beamforming vector for UE $k$ as \cite{origSINR}

\begin{singlespace}
\noindent
\begin{equation}
\mathbf{w}_{k}^{\textrm{SLNR}}=\sqrt{\tilde{P}_{k}}\boldsymbol{\textrm{v}}_{\textrm{max}}\left\{ \left(\frac{N_{0}}{\tilde{P}_{k}}\mathbf{I}_{N}+\mathbf{\mathbf{\bar{\check{H}}}_{\mathit{k}}^{\textrm{H}}\mathbf{\bar{\check{H}}}_{\mathit{k}}}\right)^{-1}\left(\xi_{k}^{2}\mathbf{\check{h}}_{k}^{\textrm{H}}\mathbf{\check{h}}_{k}\right)\right\} ,\label{eq:wk_SLNR}
\end{equation}

\end{singlespace}

\noindent where $\boldsymbol{\textrm{v}}_{\textrm{max}}\left\{ \mathbf{Q}\right\} $
denotes the eigen-vector associated with the largest eigen-value of
matrix $\mathbf{Q}$ and $\mathbf{\bar{\check{H}}}_{\mathit{k}}$
is an extended channel matrix excluding $\xi_{k}\mathbf{\mathbf{\check{h}}}_{k}$
from $\check{\mathbf{H}}$ ,i.e., $\mathbf{\bar{\check{H}}}_{\mathit{k}}=\left[\xi_{1}\mathbf{\check{h}}_{1}^{\textrm{T}},\ldots,\xi_{k-1}\mathbf{\mathbf{\check{h}}}_{k-1}^{\textrm{T}},\xi_{k+1}\mathbf{\mathbf{\check{h}}}_{k+1}^{\textrm{T}},\ldots,\xi_{K}\mathbf{\check{h}}_{K}^{\textrm{T}}\right]^{\textrm{T}}$.

\section{The Proposed Leakage-based Robust Beamforming with Per-antenna Power
Constraints}

As for robust beamforming schemes, we should take the imperfect CDI
into account when designing UEs' beamforming vectors, i.e., $\tilde{\mathbf{h}}_{k}$
will be considered as a random vector $\tilde{\mathbf{h}}_{k}^{\diamond}$
isotropically distributed around $\hat{\mathbf{h}}_{k}$. Similar
to \eqref{eq:decomp_hk_tilde}, $\tilde{\mathbf{h}}_{k}^{\diamond}$
can be written as \cite{Limited-bit CSI}

\noindent
\begin{equation}
\tilde{\mathbf{h}}_{k}^{\diamond}=\sqrt{1-Z}\mathbf{\hat{h}}_{k}+\sqrt{Z}\mathbf{e}_{k}^{\diamond},\label{eq:decomp_hk_tilde_BS}
\end{equation}

\noindent where $\mathbf{e}_{k}^{\diamond}$ is isotropically distributed
in the $\left(N-1\right)$-dimensional nullspace of $\mathbf{\hat{h}}_{k}$
and the random variable $Z$ is defined as $Z=\sin^{2}\left(\angle\left(\tilde{\mathbf{h}}_{k}^{\diamond},\mathbf{\hat{h}}_{k}\right)\right)$.

Based on the idea of controlling the interference leakage to implicitly
maximize UEs' SINRs \cite{origSINR}, recently in \cite{basic_robust_SLNR},
\cite{robust_AveSLNR}, \cite{robust_ProbLeak1} and \cite{robust_ProbLeak_ICC12},
the authors proposed new robust beamforming schemes for MU-MISO systems
under quantized CDI and per-UE power constraints. In \cite{robust_AveSLNR},
the SLNR maximization problem with regard to $\mathbf{w}_{k}$ was
transformed to a Rayleigh quotient problem with a lower bound solution
by utilizing Jensen's inequality to maximize each UE's average SLNR
shown as

\begin{singlespace}
\noindent
\begin{eqnarray}
\underset{\mathbf{w}_{k}}{\max} &  & \frac{\mathbb{E}_{\left[\tilde{\mathbf{h}}_{j}^{\diamond}\left|\hat{\mathbf{h}}_{j}\right.j=1,2,\ldots,K\right]}\left\{ S_{k}^{\diamond}\right\} }{\mathbb{E}_{\left[\tilde{\mathbf{h}}_{j}^{\diamond}\left|\hat{\mathbf{h}}_{j}\right.j=1,2,\ldots,K\right]}\left\{ L_{k}^{\diamond}\right\} +N_{0}}\nonumber \\
\textrm{s.t.} &  & \textrm{tr}\left\{ \mathbf{w}_{k}\mathbf{w}_{k}^{\textrm{H}}\right\} \leq\tilde{P}_{k},\label{eq:Problem_R_SLNRk}
\end{eqnarray}

\end{singlespace}

\noindent where the signal power $S_{k}^{\diamond}$ and leakage power
$L_{k}^{\diamond}$ are defined as $S_{k}^{\diamond}=A_{k}^{\textrm{ave}}\left|\xi_{k}\tilde{\mathbf{h}}_{k}^{\diamond}\mathbf{w}_{k}\right|^{2}$
and $L_{k}^{\diamond}=\sum_{j=1,j\neq k}^{K}A_{j}^{\textrm{ave}}\left|\xi_{j}\tilde{\mathbf{h}}_{j}^{\diamond}\mathbf{w}_{k}\right|^{2}$,
respectively. For brevity, we omit the subscription $\left[\tilde{\mathbf{h}}_{j}^{\diamond}\left|\hat{\mathbf{h}}_{j}\right.j=1,2,\ldots,K\right]$
of $\mathbb{E}$ hereafter. According to \cite{robust_AveSLNR}, $\mathbb{E}\left\{ S_{k}^{\diamond}\right\} $
can be computed as

\begin{singlespace}
\noindent
\begin{eqnarray}
\mathbb{E}\left\{ S_{k}^{\diamond}\right\}  & = & \xi_{k}^{2}A_{k}^{\textrm{ave}}\mathbf{w}_{k}^{\textrm{H}}\mathbf{U}_{k}\mathbf{w}_{k},\label{eq:E{Sk}}
\end{eqnarray}

\end{singlespace}

\noindent where $\mathbf{U}_{k}=\left(1-\frac{N\eta}{N-1}\right)\mathbf{\hat{h}}_{k}^{\textrm{H}}\mathbf{\hat{h}}_{k}+\frac{\eta}{N-1}\mathbf{I}_{N}$
and $\eta$ is computed as $\eta=2^{B}\beta\left(2^{B},\frac{N}{N-1}\right)$
\cite{Limited-bit CSI}. Here, $\beta\left(x,y\right)$ is the beta
function defined as $\beta\left(x,y\right)=\frac{\Gamma\left(x\right)\Gamma\left(y\right)}{\Gamma\left(x+y\right)}$
\cite{integral_table book}, where $\Gamma\left(\cdot\right)$ denotes
the gamma function \cite{integral_table book}. It should be noted
that $\mathbf{U}_{k}$ is positive definite because it is easy to
verify

\begin{singlespace}
\noindent
\begin{equation}
1-\frac{N\eta}{N-1}>0,\quad\textrm{for }N>1,B\geq0.\label{eq:ineq_posi_def}
\end{equation}

\end{singlespace}

\noindent Similar to \eqref{eq:E{Sk}}, $\mathbb{E}\left\{ L_{k}^{\diamond}\right\} $
can be derived as

\begin{singlespace}
\noindent
\begin{equation}
\mathbb{E}\left\{ L_{k}^{\diamond}\right\} =\mathbf{w}_{k}^{\textrm{H}}\mathbf{\bar{U}}_{k}\mathbf{w}_{k},\label{eq:E{Lk}}
\end{equation}

\end{singlespace}

\noindent where $\mathbf{\bar{U}}_{k}=\sum_{j=1,j\neq k}^{K}\xi_{j}^{2}A_{j}^{\textrm{ave}}\mathbf{U}_{j}$.
Then the solution to problem \eqref{eq:Problem_R_SLNRk} can be written
in a similar expression as in \eqref{eq:wk_SLNR} with $\mathbf{\mathbf{\bar{\check{H}}}_{\mathit{k}}^{\textrm{H}}\mathbf{\bar{\check{H}}}_{\mathit{k}}}$
and $\mathbf{\check{h}}_{k}^{\textrm{H}}\mathbf{\check{h}}_{k}$ respectively
replaced by $\mathbf{\bar{U}}_{k}$ and $A_{k}^{\textrm{ave}}\mathbf{U}_{k}$
\cite{robust_AveSLNR}. The scheme based on problem \eqref{eq:Problem_R_SLNRk}
is named the average SLNR (ASLNR) scheme by the authors of \cite{robust_AveSLNR}.

In \cite{basic_robust_SLNR}, the optimization of SLNR was interpreted
as keeping the expectation of leakage power below a threshold while
maximizing the expectation of signal power. The optimization problem
with regard to $\mathbf{w}_{k}$ can be formulated as

\begin{singlespace}
\noindent
\begin{eqnarray}
\underset{\mathbf{w}_{k}}{\max} &  & \mathbb{E}\left\{ S_{k}^{\diamond}\right\} \nonumber \\
\textrm{s.t.} &  & \mathbb{E}\left\{ L_{k}^{\diamond}\right\} \leq\gamma_{k},\textrm{ and }\textrm{tr}\left\{ \mathbf{w}_{k}\mathbf{w}_{k}^{\textrm{H}}\right\} \leq\tilde{P}_{k},\label{eq:Problem_LC}
\end{eqnarray}

\end{singlespace}

\noindent where $\gamma_{k}$ is a design parameter of the leakage
power threshold. Problem \eqref{eq:Problem_LC} is a non-convex quadratically
constrained quadratic program (QCQP) problem \cite{vec_randm}. However,
it can be transformed to an equivalent SDP problem as

\begin{singlespace}
\noindent
\begin{eqnarray}
\underset{\mathbf{Q}_{k}\in\mathbb{\mathbb{H}}_{N}^{\mathit{+}}}{\max} &  & \textrm{tr}\left\{ \mathbf{Q}_{k}\mathbf{U}_{k}\right\} \nonumber \\
\textrm{s.t.} &  & \textrm{tr}\left\{ \mathbf{Q}_{k}\mathbf{\bar{U}}_{k}\right\} \leq\gamma_{k},\textrm{ and }\textrm{tr}\left\{ \mathbf{Q}_{k}\right\} \leq\tilde{P}_{k};\nonumber \\
 &  & \textrm{rank}\left\{ \mathbf{Q}_{k}\right\} =1.\label{eq:Problem_LC_SDP}
\end{eqnarray}

\end{singlespace}

\noindent Problem \eqref{eq:Problem_LC_SDP} is still a non-convex
problem due to the non-convex rank-one constraint. Applying the SDP
relaxation technique \cite{vec_randm} by omitting the rank-one constraint,
we get

\begin{singlespace}
\noindent
\begin{eqnarray}
\underset{\mathbf{Q}_{k}\in\mathbb{\mathbb{H}}_{N}^{\mathit{+}}}{\max} &  & \textrm{tr}\left\{ \mathbf{Q}_{k}\mathbf{U}_{k}\right\} \nonumber \\
\textrm{s.t.} &  & \textrm{tr}\left\{ \mathbf{Q}_{k}\mathbf{\bar{U}}_{k}\right\} \leq\gamma_{k},\textrm{ and }\textrm{tr}\left\{ \mathbf{Q}_{k}\right\} \leq\tilde{P}_{k}.\label{eq:Problem_LC_SDPR}
\end{eqnarray}

\end{singlespace}

\noindent Now problem \eqref{eq:Problem_LC_SDPR} is a standard convex
SDP problem and can be solved efficiently using the mathematical software
package \cite{cvx_software}. According to \cite{basic_robust_SLNR},
the solution of an SDP problem like problem \eqref{eq:Problem_LC_SDPR}
is always rank-one if it has at most three constraints. Since problem
\eqref{eq:Problem_LC_SDPR} has only two constraints, we can conclude
that problem \eqref{eq:Problem_LC_SDP} and \eqref{eq:Problem_LC_SDPR}
are equivalent. Thus the solution for the original problem \eqref{eq:Problem_LC}
can be extracted from the solution to problem \eqref{eq:Problem_LC_SDPR}.

Obviously, in problem \eqref{eq:Problem_LC} $\gamma_{k}$ should
be carefully chosen to make sure that the leakage power constraints
are neither too tight nor too loose to achieve a good balance between
the maximization of $\mathbb{E}\left\{ S_{k}^{\diamond}\right\} $
and the minimization of $\mathbb{E}\left\{ L_{k}^{\diamond}\right\} $.
It is worthwhile to note that recently in \cite{robust_ProbLeak1}
and \cite{robust_ProbLeak_ICC12}, the authors proposed a probabilistic
approach in the control of interference leakage, i.e., to keep the
event of large leakage below a certain probability. The scheme will
be referred to as the probabilistic leakage control (PLC) scheme hereafter.
In the PLC scheme, the leakage constraint in problem \eqref{eq:Problem_LC}
is regulated as

\begin{singlespace}
\noindent
\begin{equation}
\Pr\left(\mathbb{E}\left\{ L_{k}^{\diamond}\right\} \geq\gamma_{k}\right)\leq p_{k},\label{eq:Prob_leakage_constraint}
\end{equation}

\end{singlespace}

\noindent where $p_{k}$ is a given probability of the event that
$\mathbb{E}\left\{ L_{k}^{\diamond}\right\} $ exceeds $\gamma_{k}$.
According to \cite{robust_ProbLeak1} and \cite{robust_ProbLeak_ICC12},
by invoking Markov's inequality, \eqref{eq:Prob_leakage_constraint}
can be nicely tightened by a new constraint $\mathbb{E}\left\{ L_{k}^{\diamond}\right\} \leq p_{k}\gamma_{k}$,
which generates a new problem shown as

\begin{singlespace}
\noindent
\begin{eqnarray}
\underset{\mathbf{w}_{k}}{\max} &  & \mathbb{E}\left\{ S_{k}^{\diamond}\right\} \nonumber \\
\textrm{s.t.} &  & \mathbb{E}\left\{ L_{k}^{\diamond}\right\} \leq p_{k}\gamma_{k},\textrm{ and }\textrm{tr}\left\{ \mathbf{w}_{k}\mathbf{w}_{k}^{\textrm{H}}\right\} \leq\tilde{P}_{k}.\label{eq:Problem_PLC}
\end{eqnarray}

\end{singlespace}

\noindent Problem \eqref{eq:Problem_PLC} is essentially equivalent
to problem \eqref{eq:Problem_LC} if we define $p_{k}\gamma_{k}$
as a new threshold $\tilde{\gamma}_{k}$. Therefore, the issue still
remains regarding the appropriate proposal of $\gamma_{k}$.

\subsection{Minimum Average Leakage Power}

\noindent In order to get some insights on the design of $\gamma_{k}$,
we will start with the comparison between the average leakage power
in problem \eqref{eq:Problem_LC} and that of the non-robust ZF beamforming
given by \eqref{eq:NR_ZF_wk}. Our result is summarized in Theorem~\ref{thm:Theorem_1}.
\begin{thm}
\label{thm:Theorem_1}Let $\mathbf{w}_{k}\left(k\neq j\right)$ be
a general beamforming vector expressed as $\mathbf{w}_{k}=\sqrt{\tilde{P}_{k}}\mathbf{\tilde{w}}_{k}$,
where $\mathbf{\tilde{w}}_{k}$ is the normalized vector of $\mathbf{w}_{k}$.
Then for arbitrary CF-CMI $A_{j}$, the expected interference leakage
from UE $k$ to $j$ is lower bounded by

\vspace*{-10pt}
\begin{eqnarray}
\mathbb{E}\left\{ \left|\xi_{j}\sqrt{A_{j}}\tilde{\mathbf{h}}_{j}^{\diamond}\mathbf{w}_{k}\right|^{2}\right\}  & \geq & \mathbb{E}\left\{ \left|\xi_{j}\sqrt{A_{j}}\tilde{\mathbf{h}}_{j}^{\diamond}\mathbf{w}_{k}^{\textrm{ZF}}\right|^{2}\right\} \nonumber \\
 & = & \frac{\tilde{P}_{k}\xi_{j}^{2}A_{j}\eta}{\left(N-1\right)}.\label{eq:Theorem_1}
\end{eqnarray}

\vspace*{-10pt}
\end{thm}
\begin{IEEEproof}
\noindent See Appendix I.
\end{IEEEproof}
\vspace{5pt}
From Theorem~\ref{thm:Theorem_1}, we can state that the non-robust
ZF beamforming scheme is optimal in the sense of generating the minimum
amount of average inter-UE interference under limited-bit CDI. In
other words, the average leakage power of the non-robust ZF beamforming
scheme can serve as a lower bound for the interference leakage in
problem \eqref{eq:Problem_LC}.

\subsection{The Proposed Minimum Average Leakage Control Beamforming with Per-UE
Power Constraints}

According to Theorem~\ref{thm:Theorem_1}, under the assumptions
of average CF-CMI and the Rayleigh channel fading, we can set the
leakage power threshold $\gamma_{k}$ to the derived lower bound as

\begin{singlespace}
\noindent
\begin{equation}
\gamma_{k}^{\textrm{MALC}}=\frac{\tilde{P}_{k}N\eta}{\left(N-1\right)}\sum_{j=1,j\neq k}^{K}\xi_{j}^{2},\label{eq:gamk_ZF_1_aveCMI}
\end{equation}

\end{singlespace}

\noindent where $A_{j}$ in Theorem~\ref{thm:Theorem_1} has been
replaced by $A_{j}^{\textrm{ave}}=N$, considering that for Rayleigh
fading channels $\left\Vert \mathbf{h}_{j}\right\Vert ^{2}$ follows
a chi-squared distribution with $2N$ degrees of freedom and its mean
is $N$ \cite{Proakis book}. On the other hand, when the BS has quantized
CF-CMI, we can substitute $A_{j}$ with $\hat{A}_{j}$ and choose
$\gamma_{k}^{\textrm{MALC}}$ as

\begin{singlespace}
\noindent
\begin{equation}
\gamma_{k}^{\textrm{MALC}}=\frac{\tilde{P}_{k}\eta}{\left(N-1\right)}\sum_{j=1,j\neq k}^{K}\xi_{j}^{2}\hat{A}_{j}.\label{eq:gamk_ZF_2_quantCMI}
\end{equation}

\end{singlespace}

\noindent Then we can re-write problem \eqref{eq:Problem_LC} as

\begin{singlespace}
\noindent
\begin{eqnarray}
\underset{\mathbf{w}_{k}}{\max} &  & \mathbb{E}\left\{ S_{k}^{\diamond}\right\} \nonumber \\
\textrm{s.t.} &  & \mathbb{E}\left\{ L_{k}^{\diamond}\right\} \leq\gamma_{k}^{\textrm{MALC}},\textrm{ and }\textrm{tr}\left\{ \mathbf{w}_{k}\mathbf{w}_{k}^{\textrm{H}}\right\} \leq\tilde{P}_{k}.\label{eq:Problem_R_ZF}
\end{eqnarray}

\end{singlespace}

\noindent The beamforming scheme based on problem \eqref{eq:Problem_R_ZF}
will be called the minimum average leakage control (MALC) scheme since
it controls the expected leakage power according to the minimum average
leakage given by Theorem~\ref{thm:Theorem_1}.

\subsection{The Proposed Relaxed Average Leakage Control Beamforming with Per-UE
Power Constraints}

From Theorem~\ref{thm:Theorem_1}, we have found the minimum threshold
value for the average interference leakage shown in \eqref{eq:gamk_ZF_1_aveCMI}/\eqref{eq:gamk_ZF_2_quantCMI}.
Next, we want to raise the leakage threshold so that the weighted
sum-rate can be increased. The intuition is that interference minimization,
e.g., ZF precoding, is generally not the optimal strategy for throughput
maximization or MSE minimization. Allowing some more interference
leakage can increase the weighted sum-rate due to signal power boosting.
Of course if the tolerated leakage is set to be too large, the rate
will inevitably decrease. Our intention is to find an appropriate
leakage level, which can generate a good weighted sum-rate performance.

Under the assumption of average CF-CMI, we propose that the upper
half segment of the CDF of $\left|\xi_{j}\sqrt{A_{j}^{\diamond}}\tilde{\mathbf{h}}_{j}^{\diamond}\mathbf{w}_{k}^{\textrm{ZF}}\right|^{2}$
should serve as the analytical reference for determining how large
$\gamma_{k}$ should be, where $A_{j}^{\diamond}$ is the randomly
reconstructed CF-CMI at the BS based on $A_{j}^{\textrm{ave}}$ and
the channel-fading model. Our proposal is based on the fact that the
CDF of $\left|\xi_{j}\sqrt{A_{j}^{\diamond}}\tilde{\mathbf{h}}_{j}^{\diamond}\mathbf{w}_{k}^{\textrm{ZF}}\right|^{2}$
contains tractable information on the average leakage level of a conservative
beamformer $\mathbf{w}_{k}^{\textrm{ZF}}$ with minimum generation
of inter-UE interference. As a result, $\gamma_{k}^{\textrm{MALC}}$
will be loosened to another threshold $\gamma_{k}^{\textrm{RALC}}$
and we can design a relaxed average leakage control (RALC) beamforming
scheme with per-UE power constraints. Since $\left|\xi_{j}\sqrt{A_{j}^{\diamond}}\tilde{\mathbf{h}}_{j}^{\diamond}\mathbf{w}_{k}^{\textrm{ZF}}\right|^{2}=\tilde{P}_{k}\xi_{j}^{2}\left|\sqrt{A_{j}^{\diamond}}\tilde{\mathbf{h}}_{j}^{\diamond}\mathbf{\tilde{w}}_{k}^{\textrm{ZF}}\right|^{2}$,
we can derive the CDF of $\left|\sqrt{A_{j}^{\diamond}}\tilde{\mathbf{h}}_{j}^{\diamond}\mathbf{\tilde{w}}_{k}^{\textrm{ZF}}\right|^{2}$
and scale it by a factor of $\tilde{P}_{k}\xi_{j}^{2}$ to obtain
the CDF of $\left|\xi_{j}\sqrt{A_{j}^{\diamond}}\tilde{\mathbf{h}}_{j}^{\diamond}\mathbf{w}_{k}^{\textrm{ZF}}\right|^{2}$.
In the following we present our result on the CDF of $\left|\sqrt{A_{j}^{\diamond}}\tilde{\mathbf{h}}_{j}^{\diamond}\mathbf{\tilde{w}}_{k}^{\textrm{ZF}}\right|^{2}$
for Rayleigh fading channels in Theorem~\ref{thm:Theorem_2}.
\begin{thm}
\label{thm:Theorem_2}Let $D=\left|\sqrt{A_{j}^{\diamond}}\tilde{\mathbf{h}}_{j}^{\diamond}\mathbf{\tilde{w}}_{k}^{\textrm{ZF}}\right|^{2}$.
Then for Rayleigh fading channels the CDF of $D$ is shown in \eqref{eq:Theorem_2},
where $\textrm{E}_{1}\left(x\right)=\int_{x}^{\infty}\frac{\textrm{e}^{-t}}{t}dt$
is the exponential integral function \textup{\cite{integral_table book}}.
\begin{figure*}[!tp]
\noindent
\begin{eqnarray}
P_{D}\left(d\right) & = & 1+\frac{\textrm{e}^{-d}}{\left(N-2\right)!}\left\{ \sum_{n=0}^{N-2}\sum_{m=1}^{2^{B}}\textrm{C}_{N-2}^{n}\textrm{C}_{2^{B}}^{m}\frac{\left(-1\right)^{n+m}m}{mN-\left(m+n\right)}\sum_{l=0}^{N-1-n}l!\textrm{C}_{N-1-n}^{l}d^{N-1-l}\right\} \nonumber \\
 &  & -\frac{\textrm{e}^{-d}}{\left(N-2\right)!}\left\{ \sum_{n=0}^{N-2}\textrm{C}_{N-2}^{n}\textrm{C}_{2^{B}}^{1}\frac{\left(-1\right)^{n+1}}{N-1-n}d^{N-1}\right\} \nonumber \\
 &  & -\frac{\textrm{e}^{-d}}{\left(N-2\right)!}\left\{ \sum_{n=0}^{N-2}\sum_{m=2}^{2^{B}}\textrm{C}_{N-2}^{n}\textrm{C}_{2^{B}}^{m}\frac{\left(-1\right)^{n+m}m}{mN-\left(m+n\right)}\left[\sum_{l=1}^{\left(m-1\right)\left(N-1\right)-1}\frac{\left(-1\right)^{l-1}d^{N-1+l}}{l!\textrm{C}_{\left(m-1\right)\left(N-1\right)-1}^{l}}\right]\right\} \nonumber \\
 &  & -\frac{\textrm{E}_{1}\left(d\right)}{\left(N-2\right)!}\left\{ \sum_{n=0}^{N-2}\sum_{m=2}^{2^{B}}\textrm{C}_{N-2}^{n}\textrm{C}_{2^{B}}^{m}\frac{\left(-1\right)^{n+m}m}{mN-\left(m+n\right)}\frac{\left(-1\right)^{\left(m-1\right)\left(N-1\right)-1}d^{m\left(N-1\right)}}{\left(\left(m-1\right)\left(N-1\right)-1\right)!}\right\} .\label{eq:Theorem_2}
\end{eqnarray}

\rule[0.5ex]{2.05\columnwidth}{1pt}
\end{figure*}
\end{thm}
\begin{IEEEproof}
\noindent See Appendix II.
\end{IEEEproof}
\vspace{5pt}
If the BS has quantized CF-CMI $\hat{A}_{j}$, then $A_{j}^{\diamond}$
should take the value of $\hat{A}_{j}$. Consequently, we can get
the CDF of $\left|\xi_{j}\sqrt{\hat{A}_{j}}\tilde{\mathbf{h}}_{j}^{\diamond}\mathbf{w}_{k}^{\textrm{ZF}}\right|^{2}$
by scaling that of $\left|\tilde{\mathbf{h}}_{j}^{\diamond}\mathbf{\tilde{w}}_{k}^{\textrm{ZF}}\right|^{2}$
by a factor of $\tilde{P}_{k}\xi_{j}^{2}\hat{A}_{j}$. In Theorem~\ref{thm:Theorem_3},
we show our result on the CDF of $\left|\tilde{\mathbf{h}}_{j}^{\diamond}\mathbf{\tilde{w}}_{k}^{\textrm{ZF}}\right|^{2}$.
\begin{thm}
\label{thm:Theorem_3}Let $V=\left|\tilde{\mathbf{h}}_{j}^{\diamond}\mathbf{\tilde{w}}_{k}^{\textrm{ZF}}\right|^{2}$.
Then the CDF of $V$ is\vspace{5pt}

\noindent $P_{V}\left(v\right)=1+\left(N-1\right)\times$

\noindent \vspace{-20pt}

\begin{equation}
\sum_{n=0}^{N-2}\sum_{m=1}^{2^{B}}\textrm{C}_{N-2}^{n}\textrm{C}_{2^{B}}^{m}\left(-1\right)^{n+m}m\frac{v^{n}-v^{m\left(N-1\right)}}{mN-\left(m+n\right)}.\label{eq:Theorem_3}
\end{equation}
\end{thm}
\begin{IEEEproof}
\noindent See Appendix III.
\end{IEEEproof}
\vspace{5pt}
Based on Theorem~\ref{thm:Theorem_2} or \ref{thm:Theorem_3}, we
can relax the control target of the interference leakage to the $\delta_{k}$
percent point of the leakage power of the non-robust ZF beamforming
scheme. To be more specific, considering average CF-CMI, we can set
$\gamma_{k}^{\textrm{RALC}}$ according to Theorem~\ref{thm:Theorem_2}
as

\begin{singlespace}
\noindent
\begin{equation}
\gamma_{k}^{\textrm{RALC}}=\tilde{P}_{k}\sum_{j=1,j\neq k}^{K}\xi_{j}^{2}P_{D}^{-1}\left(\delta_{k}\right),\label{eq:gamk_RZF_1_aveCMI}
\end{equation}

\end{singlespace}

\noindent where $P_{D}^{-1}\left(\delta_{k}\right)=\underset{d}{\arg}\left\{ P_{D}\left(d\right)=\delta_{k}\right\} $.
When quantized CF-CMI is available at the BS, we can recall Theorem~\ref{thm:Theorem_3}
and choose $\gamma_{k}^{\textrm{RALC}}$ as

\begin{singlespace}
\noindent
\begin{equation}
\gamma_{k}^{\textrm{RALC}}=\tilde{P}_{k}\sum_{j=1,j\neq k}^{K}\xi_{j}^{2}\hat{A}_{j}P_{V}^{-1}\left(\delta_{k}\right),\label{eq:gamk_RZF_2_quantCMI}
\end{equation}

\end{singlespace}

\noindent where $P_{V}^{-1}\left(\delta_{k}\right)=\underset{v}{\arg}\left\{ P_{V}\left(v\right)=\delta_{k}\right\} $.
Although there is no closed-form expression for the computation of
$P_{D}^{-1}\left(\delta_{k}\right)$ or $P_{V}^{-1}\left(\delta_{k}\right)$,
it can be conveniently found by the bisection method for a given $\delta_{k}$
since $P_{D}\left(d\right)$ or $P_{V}\left(v\right)$ is a bounded
and monotonically increasing function. In the bisection method, considering
$P_{D}\left(d\right)$ as an example, first we choose a small value
$d_{1}$ and a large one $d_{2}$ to construct an interval $\left[P_{D}\left(d_{1}\right),P_{D}\left(d_{2}\right)\right]$
that contains $\delta_{k}$, then we continually compare $P_{D}\left(\frac{d_{1}+d_{2}}{2}\right)$
with $\delta_{k}$ and update $d_{1}$ or $d_{2}$ with $\frac{d_{1}+d_{2}}{2}$
on condition that $\delta_{k}$ stays inside the interval $\left[P_{D}\left(d_{1}\right),P_{D}\left(d_{2}\right)\right]$.
The bisection searching stops when $\left[P_{D}\left(d_{1}\right),P_{D}\left(d_{2}\right)\right]$
is sufficiently narrow, and then we can obtain $\delta_{k}=\frac{d_{1}+d_{2}}{2}$.
As for the choice of $\delta_{k}$ in the proposed RALC scheme, since
we want to get a relaxed leakage threshold compared with $\gamma_{k}^{\textrm{MALC}}$,
it is intuitive to look beyond $\gamma_{k}^{\textrm{MALC}}$ and hinge
the leakage power threshold to some point on the upper half segment
of $P_{D}\left(d\right)$ or $P_{V}\left(v\right)$, i.e.,

\begin{singlespace}
\noindent
\[
\delta_{k}\in\begin{cases}
\left(P_{D}\left(\frac{\gamma_{k}^{\textrm{MALC}}}{\tilde{P}_{k}\sum_{j\neq k}\xi_{j}^{2}}\right),1\right), & \textrm{for }\gamma_{k}^{\textrm{MALC}}\textrm{ from }\eqref{eq:gamk_ZF_1_aveCMI};\\
\left(P_{V}\left(\frac{\gamma_{k}^{\textrm{MALC}}}{\tilde{P}_{k}\sum_{j\neq k}\xi_{j}^{2}\hat{A}_{j}}\right),1\right), & \textrm{for }\gamma_{k}^{\textrm{MALC}}\textrm{ from }\eqref{eq:gamk_ZF_2_quantCMI}.
\end{cases}
\]

\end{singlespace}

\noindent Based on \eqref{eq:gamk_RZF_1_aveCMI} or \eqref{eq:gamk_RZF_2_quantCMI},
the original problem \eqref{eq:Problem_LC} becomes

\begin{singlespace}
\noindent
\begin{eqnarray}
\underset{\mathbf{w}_{k}}{\max} &  & \mathbb{E}\left\{ S_{k}^{\diamond}\right\} \nonumber \\
\textrm{s.t.} &  & \mathbb{E}\left\{ L_{k}^{\diamond}\right\} \leq\gamma_{k}^{\textrm{RALC}},\textrm{ and }\textrm{tr}\left\{ \mathbf{w}_{k}\mathbf{w}_{k}^{\textrm{H}}\right\} \leq\tilde{P}_{k}.\label{eq:Problem_R_RZF}
\end{eqnarray}

\end{singlespace}

The beamforming scheme based on problem \eqref{eq:Problem_R_RZF}
will be referred to as the relaxed average leakage control (RALC)
scheme because it relaxes the leakage threshold compared with the
MALC scheme. It should be noted that though $\delta_{k}$ is a design
parameter chosen to allow more interference leakage, the leakage threshold
can be immediately found in an efficient way once the system parameters
$\left\{ N,K,B,\delta_{k}\right\} $ are provided. On the other hand,
the PLC scheme \cite{robust_ProbLeak1}, \cite{robust_ProbLeak_ICC12}
employs an empirical method to determine the leakage threshold, which
needs to exhaustively search the appropriate leakage threshold by
a great amount of simulations for each parameter set $\left\{ N,K,B\right\} $.

\subsection{The Proposed Robust Beamforming Schemes with Per-antenna Power Constraints\label{sub:The-Proposed-Scheme}}

Problems \eqref{eq:Problem_R_ZF} and \eqref{eq:Problem_R_RZF} with
the proposed leakage power thresholds are subject to power constraints
on a per-UE basis. However, the appropriate choices of $\tilde{P}_{k}$s
for those problems that can optimize the system performance measure,
e.g., the weighted sum-rate, under per-antenna power constraints are
still unclear. In this paper, we propose a two-stage algorithm to
alternately update the per-UE power allocations and beamforming vectors
in order to maximize the expected weighted sum-rate performance under
per-antenna power constraints.

Suppose that the precoding matrix in the $l$-th step is $\mathbf{W}^{\left(l\right)}=\left[\sqrt{\tilde{P}_{1}^{\left(l\right)}}\tilde{\mathbf{w}}_{1}^{\left(l\right)},\sqrt{\tilde{P}_{2}^{\left(l\right)}}\tilde{\mathbf{w}}_{2}^{\left(l\right)},\ldots,\sqrt{\tilde{P}_{K}^{\left(l\right)}}\mathbf{\tilde{w}}_{K}^{\left(l\right)}\right]$,
where $\tilde{P}_{k}^{\left(l\right)}$ and $\tilde{\mathbf{w}}_{k}^{\left(l\right)}$
are respectively the beamforming power and normalized beamforming
vector for UE $k$. Then the straightforward way to optimize the expected
weighted sum-rate is to find new $\tilde{P}_{k}^{\left(l+1\right)}$
and $\tilde{\mathbf{w}}_{k}^{\left(l+1\right)}$ that can maximize
$f\left(\mathbf{W}^{\left(l+1\right)}\right)=\mathbb{E}\left\{ \sum_{k=1}^{K}\alpha_{k}\log_{2}\left(1+\frac{S_{k}^{\diamond\left(l+1\right)}}{I_{k}^{\diamond\left(l+1\right)}+N_{0}}\right)\right\} $,
where $S_{k}^{\diamond\left(l+1\right)}$ and $I_{k}^{\diamond\left(l+1\right)}$
are computed as $S_{k}^{\diamond\left(l+1\right)}=A_{k}^{\textrm{ave}}\left|\xi_{k}\tilde{\mathbf{h}}_{k}^{\diamond}\sqrt{\tilde{P}_{k}^{\left(l+1\right)}}\tilde{\mathbf{w}}_{k}^{\left(l+1\right)}\right|^{2}$
and $I_{k}^{\diamond\left(l+1\right)}=\sum_{j=1,j\neq k}^{K}A_{k}^{\textrm{ave}}\left|\xi_{k}\tilde{\mathbf{h}}_{k}^{\diamond}\sqrt{\tilde{P}_{j}^{\left(l+1\right)}}\tilde{\mathbf{w}}_{j}^{\left(l+1\right)}\right|^{2}$,
respectively.

\vspace*{5mm}

\noindent \textit{1) Updating Per-UE Powers}

First, we concentrate on the update of $\mathbf{\widetilde{P}}^{\left(l+1\right)}=\left(\tilde{P}_{1}^{\left(l+1\right)},\tilde{P}_{2}^{\left(l+1\right)},\ldots,\tilde{P}_{K}^{\left(l+1\right)}\right)$
with the objective to maximize $f\left(\mathbf{W}^{\left(l+1\right)}\right)$
on condition of $\tilde{\mathbf{w}}_{k}^{\left(l+1\right)}=\tilde{\mathbf{w}}_{k}^{\left(l\right)}$.
However, $f\left(\mathbf{W}^{\left(l+1\right)}\right)$ is hard to
handle because it has no explicit expression. Instead, we treat $\tilde{f}\left(\mathbf{W}^{\left(l+1\right)}\right)\left|_{\tilde{\mathbf{w}}_{k}^{\left(l+1\right)}=\tilde{\mathbf{w}}_{k}^{\left(l\right)}}\right.$,
where $\tilde{f}\left(\mathbf{W}^{\left(l+1\right)}\right)=\sum_{k=1}^{K}\alpha_{k}\log_{2}\left(1+\frac{\mathbb{E}\left\{ S_{k}^{\diamond\left(l+1\right)}\right\} }{\mathbb{E}\left\{ I_{k}^{\diamond\left(l+1\right)}\right\} +N_{0}}\right)$,
as an approximated metric of the weighted sum-rate. In a similar way
as in \eqref{eq:E{Sk}}, we can obtain

\begin{equation}
\mathbb{E}\left\{ S_{k}^{\diamond\left(l+1\right)}\left|_{\tilde{\mathbf{w}}_{k}^{\left(l+1\right)}=\tilde{\mathbf{w}}_{k}^{\left(l\right)}}\right.\right\} =\tilde{P}_{k}^{\left(l+1\right)}\lambda_{k,k}^{\left(l\right)},\label{eq:E{Sk(l+1)}}
\end{equation}

\noindent where $\lambda_{k,k}^{\left(l\right)}=\xi_{k}^{2}A_{k}^{\textrm{ave}}\tilde{\mathbf{w}}_{k}^{\left(l\right)\textrm{H}}\mathbf{U}_{k}\tilde{\mathbf{w}}_{k}^{\left(l\right)}$.
Also we can get

\begin{equation}
\mathbb{E}\left\{ I_{k}^{\diamond\left(l+1\right)}\left|_{\tilde{\mathbf{w}}_{k}^{\left(l+1\right)}=\tilde{\mathbf{w}}_{k}^{\left(l\right)}}\right.\right\} =\sum_{j=1,j\neq k}^{K}\tilde{P}_{j}^{\left(l+1\right)}\lambda_{j,k}^{\left(l\right)},\label{eq:E{Ik(l)}}
\end{equation}

\noindent where $\lambda_{j,k}^{\left(l\right)}=\xi_{k}^{2}A_{k}^{\textrm{ave}}\tilde{\mathbf{w}}_{j}^{\left(l\right)\textrm{H}}\mathbf{U}_{k}\tilde{\mathbf{w}}_{j}^{\left(l\right)}$.
Note that $\lambda_{j,k}^{\left(l\right)}>0$ because of \eqref{eq:ineq_posi_def}.
Then with some mathematical manipulation, the optimization problem
to maximize $\tilde{f}\left(\mathbf{W}^{\left(l+1\right)}\right)\left|_{\tilde{\mathbf{w}}_{k}^{\left(l+1\right)}=\tilde{\mathbf{w}}_{k}^{\left(l\right)}}\right.$
with leakage control and per-antenna power constraints can be formulated
as

\begin{singlespace}
\noindent
\begin{eqnarray}
\underset{\mathbf{\widetilde{P}}^{\left(l+1\right)}}{\min} &  & \prod_{k=1}^{K}\left(\frac{\sum_{j=1,j\neq k}^{K}\tilde{P}_{j}^{\left(l+1\right)}\lambda_{j,k}^{\left(l\right)}+N_{0}}{\sum_{j=1}^{K}\tilde{P}_{j}^{\left(l+1\right)}\lambda_{j,k}^{\left(l\right)}+N_{0}}\right)^{\alpha_{k}}\nonumber \\
\textrm{s.t.} &  & \mathbb{E}\left\{ L_{k}^{\diamond\left(l+1\right)}\left|_{\tilde{\mathbf{w}}_{k}^{\left(l+1\right)}=\tilde{\mathbf{w}}_{k}^{\left(l\right)}}\right.\right\} \leq\gamma_{k}^{\left(l\right)},\quad\forall k;\nonumber \\
 &  & \left[\sum_{k=1}^{K}\tilde{P}_{k}^{\left(l+1\right)}\tilde{\mathbf{w}}_{k}^{\left(l\right)}\tilde{\mathbf{w}}_{k}^{\left(l\right)\textrm{H}}\right]_{n,n}\leq P_{n},\quad\forall n,\nonumber \\
 &  & \tilde{P}_{k}^{\left(l+1\right)}\geq0,\quad\forall k,\label{eq:Problem_power_update_GP_nonstand}
\end{eqnarray}

\end{singlespace}

\noindent where $\mathbb{E}\left\{ L_{k}^{\diamond\left(l+1\right)}\left|_{\tilde{\mathbf{w}}_{k}^{\left(l+1\right)}=\tilde{\mathbf{w}}_{k}^{\left(l\right)}}\right.\right\} =\sum_{j=1,j\neq k}^{K}\tilde{P}_{k}^{\left(l+1\right)}\lambda_{k,j}^{\left(l\right)}$
and $\gamma_{k}^{\left(l\right)}$ is the proposed leakage power threshold
in the $l$-th step and it can be updated based on $\gamma_{k}^{\textrm{MALC}}$
or $\gamma_{k}^{\textrm{RALC}}$ with $\tilde{P}_{k}$ replaced by
$\tilde{P}_{k}^{\left(l\right)}$ in \eqref{eq:gamk_ZF_1_aveCMI}/\eqref{eq:gamk_ZF_2_quantCMI}
or in \eqref{eq:gamk_RZF_1_aveCMI}/\eqref{eq:gamk_RZF_2_quantCMI}.
Besides, the second set of constraints reflects the per-antenna power
control requirement. Unfortunately, problem \eqref{eq:Problem_power_update_GP_nonstand}
is non-convex, and searching of the global optimal solution is of
high-complexity. To our best knowledge, till now geometric programming
(GP) is one of the most efficient methods to solve this specific power
allocation problem \cite{GP_problem}. According to \cite{GP_problem},
the denominator of the objective function in problem \eqref{eq:Problem_power_update_GP_nonstand}
can be lower-bounded by the geometric inequality shown as

\begin{equation}
\prod_{k=1}^{K}\left(\sum_{j=1}^{K}\tilde{P}_{j}^{\left(l+1\right)}\lambda_{j,k}^{\left(l\right)}+N_{0}\right)^{\alpha_{k}}\geq\prod_{k=1}^{K}\prod_{j=0}^{K}\left(\frac{m_{j,k}}{\mu_{j,k}}\right)^{\mu_{j,k}\alpha_{k}},\label{eq:ineq_GP_LB}
\end{equation}

\noindent where $m_{j,k}=\begin{cases}
N_{0}, & j=0\\
\tilde{P}_{j}^{\left(l+1\right)}\lambda_{j,k}^{\left(l\right)}, & j=1,2,\ldots,K
\end{cases}$ and $\mu_{j,k}=\frac{m_{j,k}}{\sum_{j=1}^{K}m_{j,k}}.$ Substituting
the denominator of the objective function in problem \eqref{eq:Problem_power_update_GP_nonstand}
with the right-hand side in \eqref{eq:ineq_GP_LB}, we can obtain
a standard GP problem formulated as

\begin{singlespace}
\noindent
\begin{eqnarray}
\underset{\mathbf{\widetilde{P}}^{\left(l+1\right)}}{\min} &  & \frac{\prod_{k=1}^{K}\left(\sum_{j=0,j\neq k}^{K}m_{j,k}\right)^{\alpha_{k}}}{\prod_{k=1}^{K}\prod_{j=0}^{K}\left(\frac{m_{j,k}}{\mu_{j,k}}\right)^{\mu_{j,k}\alpha_{k}}}\nonumber \\
\textrm{s.t.} &  & \frac{\mathbb{E}\left\{ L_{k}^{\diamond\left(l+1\right)}\left|_{\tilde{\mathbf{w}}_{k}^{\left(l+1\right)}=\tilde{\mathbf{w}}_{k}^{\left(l\right)}}\right.\right\} }{\gamma_{k}^{\left(l\right)}}\leq1,\quad\forall k;\nonumber \\
 &  & \frac{\left[\sum_{k=1}^{K}\tilde{P}_{k}^{\left(l+1\right)}\tilde{\mathbf{w}}_{k}^{\left(l\right)}\tilde{\mathbf{w}}_{k}^{\left(l\right)\textrm{H}}\right]_{n,n}}{P_{n}}\leq1,\quad\forall n,\nonumber \\
 &  & \tilde{P}_{k}^{\left(l+1\right)}\geq0,\quad\forall k.\label{eq:Problem_power_update_GP_stand}
\end{eqnarray}

\end{singlespace}

\noindent Problem \eqref{eq:Problem_power_update_GP_stand} can be
solved iteratively by using the mathematical software \cite{cvx_software}
to obtain $\mathbf{\widetilde{P}}^{\left(l+1\right)}$ followed by
updating $\left\{ \mu_{j,k}\right\} $ to form the new optimization
problem \eqref{eq:Problem_power_update_GP_stand} regarding $\mathbf{\widetilde{P}}^{\left(l+1\right)}$
\cite{GP_problem}. The iteration is terminated when the power difference
metric $PD\textrm{\_}metric=\frac{\left|\mathbf{\widetilde{P}}_{\left(i\right)}^{\left(l+1\right)}-\mathbf{\widetilde{P}}_{\left(i-1\right)}^{\left(l+1\right)}\right|}{\left|\mathbf{\widetilde{P}}_{\left(i-1\right)}^{\left(l+1\right)}\right|}$
falls below a pre-determined threshold $\epsilon$, where $i$ is
the iteration index in the GP algorithm. Note that according to \cite{GP_problem},
the GP algorithm can handle problem \eqref{eq:Problem_power_update_GP_stand}
very efficiently and output a nearly-optimal solution to the original
problem \eqref{eq:Problem_power_update_GP_nonstand}.

\vspace*{5mm}

\noindent \textit{2) Updating Per-UE Beamforming Vectors}

Next, we fix $\mathbf{\widetilde{P}}^{\left(l+1\right)}$ and update
$\left\{ \mathbf{w}_{k}^{\left(l+1\right)}\right\} $. Here, we propose
to add per-antenna power constraints into problem \eqref{eq:Problem_R_ZF}
and \eqref{eq:Problem_R_RZF}, and optimize the beamforming vectors
in an order according to the order of $\tilde{P}_{k}^{\left(l+1\right)}$.
To be more specific, suppose that $\left\{ \tilde{P}_{k}^{\left(l+1\right)}\right\} $
is arranged in a descending order as $\tilde{P}_{\pi\left(1\right)}^{\left(l+1\right)}\geq\tilde{P}_{\pi\left(2\right)}^{\left(l+1\right)}\geq\ldots\geq\tilde{P}_{\pi\left(K\right)}^{\left(l+1\right)}$,
then $\mathbf{w}_{\pi\left(k\right)}^{\left(l+1\right)}$ shall be
the $k$-th beamforming vector for optimization and the power of the
$n$-th antenna should be lower than the per-antenna power headroom
left by the previous $k-1$ UEs, which is

\begin{singlespace}
\noindent
\begin{equation}
u_{n,\pi\left(k\right)}^{\left(l+1\right)}=P_{n}-\sum_{i=1}^{k-1}\left[\mathbf{w}_{\pi\left(i\right)}^{\left(l+1\right)}\mathbf{w}_{\pi\left(i\right)}^{\left(l+1\right)\textrm{H}}\right]_{n,n}.\label{eq:power_headroom_update}
\end{equation}

\end{singlespace}

\noindent The philosophy behind our design is that, a larger $\tilde{P}_{k}^{\left(l+1\right)}$
generally indicates greater significance of treating the beamforming
vector of UE $k$ in the maximization of the weighed sum-rate, and
hence $\mathbf{w}_{k}^{\left(l+1\right)}$ should be optimized sequentially
according to the order $\pi\left(k\right)$. Based on the above discussion,
we can re-formulate problem \eqref{eq:Problem_R_ZF} and \eqref{eq:Problem_R_RZF}
as

\begin{singlespace}
\noindent
\begin{eqnarray}
\underset{\mathbf{w}_{\pi\left(k\right)}^{\left(l+1\right)}}{\max} &  & \mathbb{E}\left\{ S_{\pi\left(k\right)}^{\diamond\left(l+1\right)}\right\} \nonumber \\
\textrm{s.t.} &  & \mathbb{E}\left\{ L_{\pi\left(k\right)}^{\diamond\left(l+1\right)}\right\} \leq\gamma_{\pi\left(k\right)}^{\left(l+1\right)};\nonumber \\
 &  & \textrm{tr}\left\{ \mathbf{w}_{\pi\left(k\right)}^{\left(l+1\right)}\mathbf{w}_{\pi\left(k\right)}^{\left(l+1\right)\textrm{H}}\right\} \leq\tilde{P}_{\pi\left(k\right)}^{\left(l+1\right)};\nonumber \\
 &  & \left[\mathbf{w}_{\pi\left(k\right)}^{\left(l+1\right)}\mathbf{w}_{\pi\left(k\right)}^{\left(l+1\right)\textrm{H}}\right]_{n,n}\leq u_{n,\pi\left(k\right)}^{\left(l+1\right)},\quad\forall n,\label{eq:Problem_LC_PA}
\end{eqnarray}

\end{singlespace}

\noindent where $\gamma_{\pi\left(k\right)}^{\left(l+1\right)}$ is
computed in a similar way as $\gamma_{k}^{\left(l\right)}$ with $\tilde{P}_{k}^{\left(l\right)}$
replaced by $\tilde{P}_{\pi\left(k\right)}^{\left(l+1\right)}$. In
order to solve problem \eqref{eq:Problem_LC_PA}, we transform it
to an SDP problem similar to problem \eqref{eq:Problem_LC_SDP}, and
then we apply the SDP relaxation technique \cite{vec_randm} by dropping
the rank-one constraints so as to get the following convex SDP problem.

\begin{singlespace}
\noindent
\begin{eqnarray}
\underset{\mathbf{Q}_{\pi\left(k\right)}^{\left(l+1\right)}\in\mathbb{\mathbb{H}}_{N}^{\mathit{+}}}{\max} &  & \textrm{tr}\left\{ \mathbf{Q}_{\pi\left(k\right)}^{\left(l+1\right)}\mathbf{U}_{\pi\left(k\right)}\right\} \nonumber \\
\textrm{s.t.} &  & \textrm{tr}\left\{ \mathbf{Q}_{\pi\left(k\right)}^{\left(l+1\right)}\bar{\mathbf{U}}_{\pi\left(k\right)}\right\} \leq\gamma_{\pi\left(k\right)}^{\left(l+1\right)};\nonumber \\
 &  & \textrm{tr}\left\{ \mathbf{Q}_{\pi\left(k\right)}^{\left(l+1\right)}\right\} \leq\tilde{P}_{\pi\left(k\right)}^{\left(l+1\right)};\nonumber \\
 &  & \left[\mathbf{Q}_{\pi\left(k\right)}^{\left(l+1\right)}\right]_{n,n}\leq u_{n,\pi\left(k\right)}^{\left(l+1\right)},\quad\forall n.\label{eq:Problem_LC_PA_SDPR}
\end{eqnarray}

\end{singlespace}

\noindent Problem \eqref{eq:Problem_LC_PA_SDPR} has $N+2$ constraints,
which is larger than three for meaningful cases with $N>1$. Thereby,
it is not guaranteed that the solution $\mathbf{Q}_{\pi\left(k\right)}^{\left(l+1\right)}$
is always rank-one \cite{basic_robust_SLNR}. Here we resort to the
randomization technique \cite{vec_randm} and obtain an approximate
solution with rank-one matrix. Denote the solution to problem \eqref{eq:Problem_LC_PA_SDPR}
as $\mathbf{Q}_{\pi\left(k\right)}^{\left(l+1\right)\textrm{*}}$
and suppose that $\textrm{rank}\left\{ \mathbf{Q}_{\pi\left(k\right)}^{\left(l+1\right)\textrm{*}}\right\} >1$,
then we generate a random vector $\tilde{\mathbf{q}}_{\pi\left(k\right)}^{\left(l+1\right)}\sim\mathcal{N}\left(\mathbf{0},\mathbf{Q}_{\pi\left(k\right)}^{\left(l+1\right)\textrm{*}}\right)$
and scale it by a factor $\rho$ to make sure that all the constraints
in problem \eqref{eq:Problem_LC_PA_SDPR} are satisfied, i.e., $\mathbf{q}_{\pi\left(k\right)}^{\left(l+1\right)}=\rho\tilde{\mathbf{q}}_{\pi\left(k\right)}^{\left(l+1\right)}.$
The vector randomization process is repeated by $L_{\textrm{rand}}$
times and we select the vector with the largest performance measure
for problem \eqref{eq:Problem_LC_PA_SDPR} as the solution to problem
\eqref{eq:Problem_LC_PA}, i.e.,

\begin{equation}
\mathbf{w}_{\pi\left(k\right)}^{\left(l+1\right)}=\underset{\mathbf{q}_{\pi\left(k\right)}^{\left(l+1\right),\left(i\right)},i\in\left\{ 1,2,\ldots,L_{\textrm{rand}}\right\} }{\arg\max}\left(\textrm{tr}\left\{ \mathbf{Q}_{\pi\left(k\right)}^{\left(l+1\right),\left(i\right)}\mathbf{U}_{\pi\left(k\right)}\right\} \right),\label{eq:wk_update}
\end{equation}

\noindent where $\mathbf{Q}_{\pi\left(k\right)}^{\left(l+1\right),\left(i\right)}=\mathbf{q}_{\pi\left(k\right)}^{\left(l+1\right),\left(i\right)}\left(\mathbf{q}_{\pi\left(k\right)}^{\left(l+1\right),\left(i\right)}\right)^{\textrm{H}}$.

\vspace*{5mm}

\noindent \textit{3) The Proposed Iterative Algorithm}

With the updated $\mathbf{w}_{\pi\left(k\right)}^{\left(l+1\right)}$,
we can compute $\tilde{f}\left(\mathbf{W}^{\left(l+1\right)}\right)$
as the performance measure for the $\left(l+1\right)$-th step. In
order to find a locally optimal solution, the two-stage algorithm
can be iteratively operated with $L_{\textrm{algo1}}$ times and the
beamforming vectors associated with the largest performance measure
will be output as the final solutions. The proposed robust beamforming
with per-antenna power constraints based on threshold $\gamma_{k}^{\textrm{MALC}}$
or $\gamma_{k}^{\textrm{RALC}}$ will be respectively referred to
as the MALC-PA or RALC-PA scheme in the following, which is summarized
in Algorithm~\ref{alg:algo_MALC-RALC}. In Algorithm~\ref{alg:algo_MALC-RALC},
we initialize the beamforming vectors as those of the ZF-PA scheme
for simplicity.

\textit{}
\begin{algorithm}[H]
\noindent \begin{raggedright}
\textit{\caption{\label{alg:algo_MALC-RALC}The MALC-PA or RALC-PA scheme}
}
\par\end{raggedright}

$\textrm{{\color{white}xx}}$\textbf{\textcolor{black}{Result}}: $\left\{ \mathbf{w}_{k}^{\textrm{algo1}}\right\} $

\noindent \textbf{\textcolor{black}{\scriptsize 1}}$\textrm{{\color{white}x}}$Initialize
$\mathbf{w}_{k}^{\textrm{algo1}}=\mathbf{w}_{k}^{\textrm{ZF-PA}}$
$\left(k\in\left\{ 1,2,\ldots,K\right\} \right)$ using \eqref{eq:wk_ZF-PA};\\
\textbf{\textcolor{black}{\scriptsize 2}}$\textrm{{\color{white}x}}$Obtain
$\tilde{P}_{k}^{\left(0\right)}$ and $\tilde{\mathbf{w}}_{k}^{\left(0\right)}$
based on $\mathbf{w}_{k}^{\textrm{algo1}}$;\\
\textbf{\textcolor{black}{\scriptsize 3}}$\textrm{{\color{white}x}}$Set
$perf\textrm{\_}metric=0$;\\
\textbf{\textcolor{black}{\scriptsize 4}}$\textrm{{\color{white}x}}$\textbf{\textcolor{black}{for}}
$l:=0$ to \textit{\emph{$L_{\textrm{algo1}}$ }}\textbf{\textit{\textcolor{black}{\emph{do}}}}\\
\textbf{\textcolor{black}{\scriptsize 5}}$\textrm{{\color{white}xxx}}$Fix
$\tilde{\mathbf{w}}_{k}^{\left(l\right)}$ and update $\tilde{P}_{k}^{\left(l+1\right)}$
by solving problem \eqref{eq:Problem_power_update_GP_stand};\\
\textbf{\textcolor{black}{\scriptsize 6}}$\textrm{{\color{white}xxx}}$Sort
$\tilde{P}_{k}^{\left(l+1\right)}$ in a descendant order $\pi\left(k\right)$;\\
\textbf{\textcolor{black}{\scriptsize 7}}$\textrm{{\color{white}xxx}}$Fix
$\tilde{P}_{k}^{\left(l+1\right)}$ and update $\mathbf{w}_{\pi\left(k\right)}^{\left(l+1\right)}$
by solving problem \eqref{eq:Problem_LC_PA_SDPR};\\
\textbf{\textcolor{black}{\scriptsize 8}}$\textrm{{\color{white}xxx}}$Compute
the performance measure $\tilde{f}\left(\mathbf{W}^{\left(l+1\right)}\right)$;\\
\textbf{\textcolor{black}{\scriptsize 9}}$\textrm{{\color{white}xxx}}$\textbf{\textcolor{black}{If}}
$\tilde{f}\left(\mathbf{W}^{\left(l+1\right)}\right)>perf\textrm{\_}metric$
\textbf{\textcolor{black}{then}}\\
\textbf{\textcolor{black}{\scriptsize 10}}$\textrm{{\color{white}xxx}}$$\textrm{{\color{white}xxx}}\textrm{Update}$
$perf\textrm{\_}metric=\tilde{f}\left(\mathbf{W}^{\left(l+1\right)}\right)$;\\
\textbf{\textcolor{black}{\scriptsize 11}}$\textrm{{\color{white}xxx}}$$\textrm{{\color{white}xxx}}\textrm{Update}$
$\mathbf{w}_{k}^{\textrm{algo1}}=\mathbf{w}_{k}^{\left(l+1\right)}$,
$\left(k\in\left\{ 1,2,\ldots,K\right\} \right)$;\\
\textbf{\textcolor{black}{\scriptsize 12}} $l=l+1$;
\end{algorithm}

\section{Simulation Results and Discussions}

In this section, we present simulation results to compare the performance
of the interested beamforming schemes. In our simulations, the system
parameters are configured as $\left(N,K\right)=\left(4,4\right)$,
$\left\{ \alpha_{k}\right\} =\left[\frac{3}{2},\frac{3}{2},1,1\right]$,
$\xi_{k}=1$ and $P_{k}=\frac{P}{N}$. In the proposed MALC-PA and
RALC-PA schemes, $\delta_{k}$ equals to 0.8 for the computation of
$\gamma_{k}^{\textrm{MALC}}$ and $\gamma_{k}^{\textrm{RALC}}$. Besides,
$L_{\textrm{rand}}=1000$, $\epsilon=0.01$ and $L_{\textrm{algo1}}=1\sim10$.
Moreover, we define the BS's SNR as $SNR=P/N_{0}$. All channels are
assumed to experience uncorrelated Rayleigh fading and the entries
of $\mathbf{h}_{k}$ are i.i.d. ZMCSCG random variables with unit
variance. The results are averaged over 5,000 independent channel
realizations.

\subsection{Verifications of Theorem 2 and 3}

\begin{figure}[H]
\centering\includegraphics[width=7.8cm]{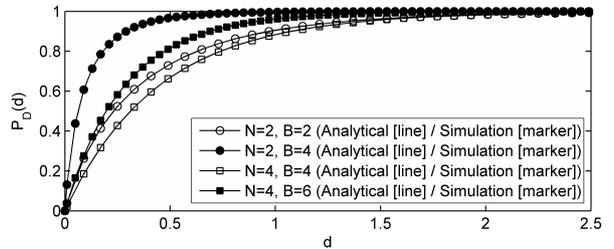} 
\vspace{-0.5em}
\renewcommand{\figurename}{Fig.}\caption{\label{fig:verification_P(d)}Simulation and analytical results of
$P_{D}\left(d\right)$ with different $\left(N,B\right)$.}
\end{figure}

\begin{figure}[H]
\centering\includegraphics[width=7.8cm]{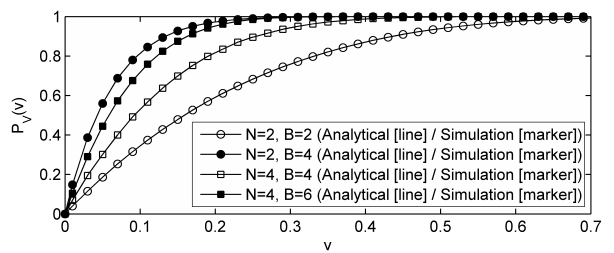} 
\vspace{-0.5em}
\renewcommand{\figurename}{Fig.}\caption{\label{fig:verification_P(v)}Simulation and analytical results of
$P_{V}\left(v\right)$ with different $\left(N,B\right)$.}
\end{figure}

Before discussing the performance of various beamforming schemes,
in Fig.~\ref{fig:verification_P(d)} and \ref{fig:verification_P(v)}
we respectively compare the numerical results of $P_{D}\left(d\right)$
and $P_{V}\left(v\right)$ with our analytical results in Theorem~\ref{thm:Theorem_2}
and \ref{thm:Theorem_3}, when $\left(N,B\right)=\left(2,2\right)$,
$\left(2,4\right)$, $\left(4,4\right)$ or $\left(4,6\right)$. As
can be seen from Fig.~\ref{fig:verification_P(d)} and \ref{fig:verification_P(v)},
the simulation results perfectly agree with our analysis, which provides
theoretical foundations for the parameter configuration of $\gamma_{k}^{\textrm{MALC}}$
and $\gamma_{k}^{\textrm{RALC}}$ in the proposed MALC-PA and RALC-PA
schemes.

\noindent

\noindent

\subsection{Average Weighted Sum-rate Performance with Per-UE Power Constraints}

\begin{figure}[H]
\centering\includegraphics[width=7.8cm]{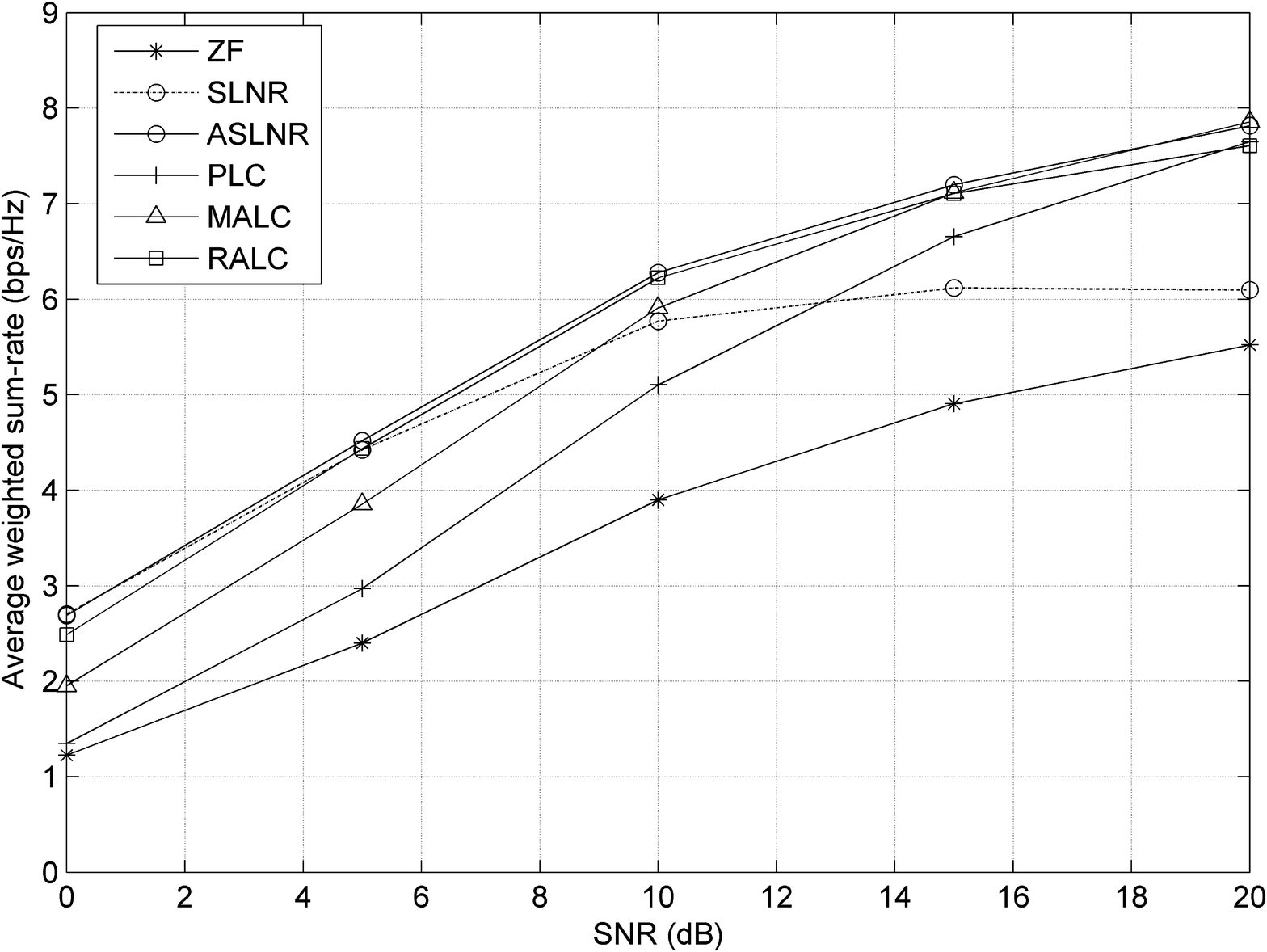} 
\vspace{-0.5em}
\renewcommand{\figurename}{Fig.}\caption{\label{fig:Perf_perUEPC}Average weighted sum-rate performance with
per-UE power constraints.}
\end{figure}

In this sub-section, we compare the average weighted sum-rate performance
of the ZF, SLNR, ASLNR, PLC schemes, together with the proposed MALC
and RALC schemes since they are all based on per-UE power constraints.
For simplicity, equal power allocation among UEs i.e., $\tilde{P}_{k}=\frac{P}{K}$,
is employed. Besides, we assume average CF-CMI and 6-bit CDI. For
the PLC scheme, $\gamma_{k}$ and $p_{k}$ are respectively set to
0.9 and 0.05 as in \cite{robust_ProbLeak_ICC12}. The performance
is exhibited in Fig.~\ref{fig:Perf_perUEPC}, from which we can see
that the ZF scheme gives the performance lower bound and the SLNR
scheme performs rather poorly in high SNR regime while the ASLNR scheme
manages to recover a large portion of the performance loss for the
SLNR scheme. Though the PLC scheme exhibits comparable performance
with the proposed MALC scheme when the SNR is high, the proposed MALC
and RALC schemes significantly outperform the PLC scheme in low to
medium SNR regimes, which shows the great significance of appropriate
choices of the leakage threshold. It is interesting to note that the
proposed MALC scheme performs better than the proposed RALC scheme
in high SNR regime since inter-UE interference dominates the performance
when the SNR is large and the strategy of minimizing the average leakage
prevails. The opposite observation can be drawn for low SNR regime.
It should also be noted that although the proposed MALC and RALC schemes
only show comparable performance with the ASLNR scheme respectively
in high and low SNR regimes, they are more flexible in handling the
per-antenna power constraints, which leads to the proposed MALC-PA
and RALC-PA schemes to be evaluated in the following sub-sections.

\noindent

\subsection{Impact of CF-CMI quantization on the Performance}

\begin{figure}[H]
\centering\includegraphics[width=7.8cm]{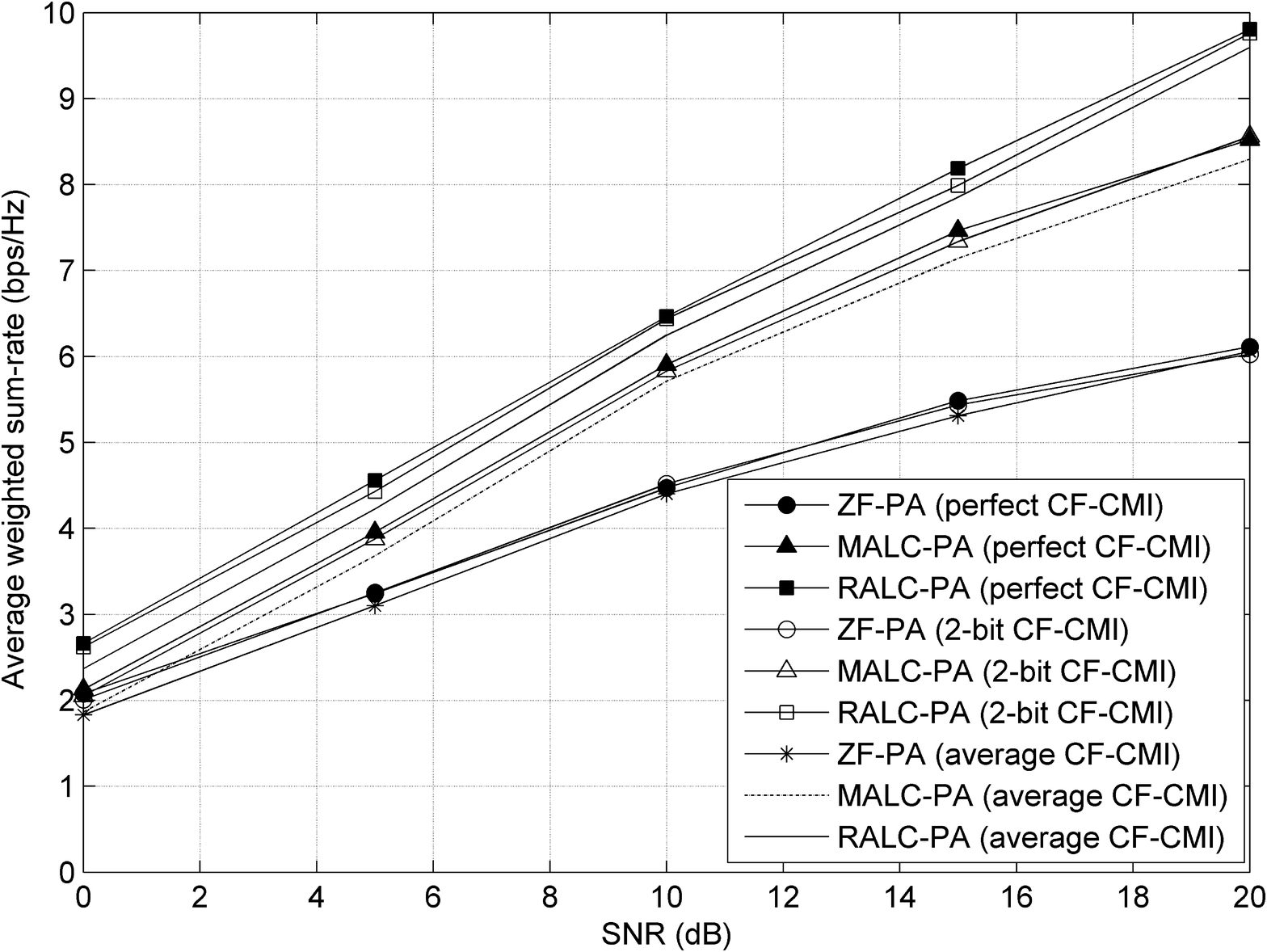} 
\vspace{-0.5em}
\renewcommand{\figurename}{Fig.}\caption{\label{fig:weighted_sumrate_diffCMI}Performance comparison with different
bits of CF-CMI quantization.}
\end{figure}

The average weighted sum-rate performance with different bits of CF-CMI
quantization is provided in Fig.~\ref{fig:weighted_sumrate_diffCMI}
for the ZF-PA scheme and the proposed MALC-PA and RALC-PA schemes
to show the impact of CF-CMI quantization on the system performance.
We assume $L_{\textrm{algo1}}=10$, 6-bit CDI and average/2-bit/perfect
CF-CMI in our simulations. For the perfect CF-CMI case, $A_{k}^{\textrm{ave}}$
is replaced by $\left\Vert \mathbf{h}_{k}\right\Vert ^{2}$ in corresponding
formulations. As can be seen from Fig.~\ref{fig:weighted_sumrate_diffCMI},
the performance curves of the average CF-CMI are very close to those
of the perfect CF-CMI. Besides, the 2-bit quantized CF-CMI is sufficient
to achieve almost the same performance as the perfect CF-CMI. Considering
the minor effectiveness of quantizing the CF-CMI on the performance,
we only consider the average CF-CMI case in the following simulations.

\noindent

\subsection{Convergence of the Proposed Algorithm}

\begin{figure}[H]
\centering\includegraphics[width=7.8cm]{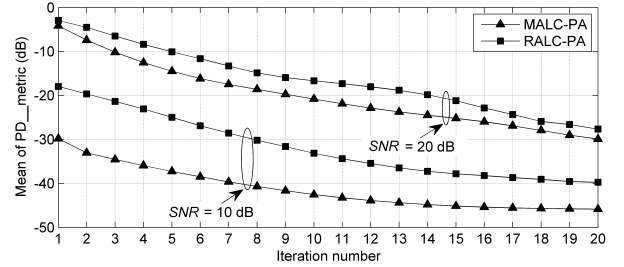} 
\vspace{-0.5em}
\renewcommand{\figurename}{Fig.}\caption{\label{fig:GP Convergence}Convergence of the power updating based
on GP.}
\end{figure}

In this sub-section, we investigate the convergence behavior of the
proposed Algorithm~\ref{alg:algo_MALC-RALC}. We assume average CF-CMI,
6-bit CDI and $SNR=$ 10 or 20 dB. First, we check the convergence
of the power updating based on the GP algorithm addressed in sub-section~\ref{sub:The-Proposed-Scheme}.
The mean value of $PD\textrm{\_}metric$ when $L_{\textrm{algo1}}=1$
is plotted in dB scale in Fig.~\ref{fig:GP Convergence}, from which
we can find that the per-UE power allocation coverges rapidly, e.g.,
only 5\textasciitilde{}15 iterations are needed for $PD\textrm{\_}metric<\epsilon$
depending on the working SNR.

\noindent

\begin{figure}[H]
\centering\includegraphics[width=7.8cm]{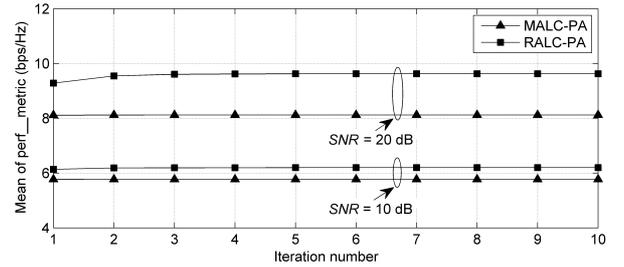} 
\vspace{-0.5em}
\renewcommand{\figurename}{Fig.}\caption{\label{fig:Algorithm Convergence}Convergence of the Algorithm~\ref{alg:algo_MALC-RALC}.}
\end{figure}

Next, we investigate the convergence of the proposed Algorithm~\ref{alg:algo_MALC-RALC}.
In Fig.~\ref{fig:Algorithm Convergence}, we plot the mean value
of $perf\textrm{\_}metric$ defined in Algorithm~\ref{alg:algo_MALC-RALC}
when $L_{\textrm{algo1}}\in\left\{ 1,2,\ldots,10\right\} $. As can
be seen from Fig.~\ref{fig:Algorithm Convergence}, both the MALC-PA
and RALC-PA schemes quickly converge to their final solutions after
merely 2 or 3 iterations, which makes the proposed beamforming schemes
feasible for practical uses.

\noindent

\subsection{Average Weighted Sum-rate Performance with Per-antenna Power Constraints}

\begin{figure}[H]
\centering\includegraphics[width=7.8cm]{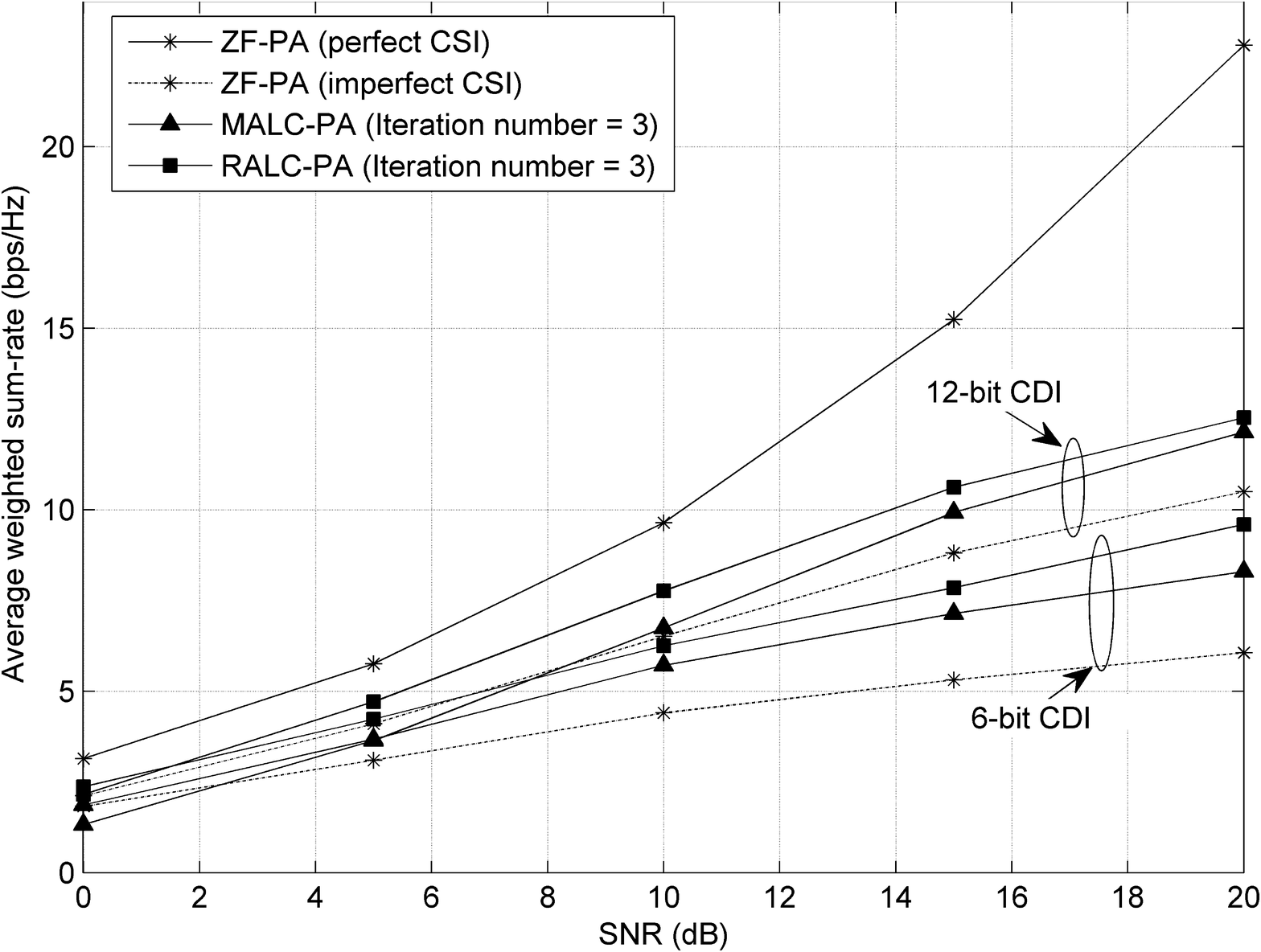} 
\vspace{-0.5em}
\renewcommand{\figurename}{Fig.}\caption{\label{fig:weighted_sumrate_averageCMI}Average weighted sum-rate
performance with per-antenna power constraints.}
\end{figure}

In Fig.~\ref{fig:weighted_sumrate_averageCMI}, we show the average
weighted sum-rate performance with per-antenna power constraints of
the proposed schemes with $L_{\textrm{algo1}}=3$ as well as the ZF-PA
scheme for the cases of average CF-CMI and 6/12-bit CDI. To show the
upper bound of the performance, we also plot the performance of the
ZF-PA scheme with perfect CSI. As can be seen from Fig.~\ref{fig:weighted_sumrate_averageCMI},
the proposed MALC-PA and RALC-PA schemes achieve considerably larger
weighted sum-rate than the ZF-PA scheme when the SNR is medium to
high, and their performance approaches that of the ZF-PA scheme with
perfect CSI at a relatively fast pace. It should be noted that the
RALC-PA scheme shows its advantage over the MALC-PA scheme in all
SNR regimes. This is because that the interference minimization strategy
is generally not an optimal one and tolerating an appropriate amount
of inter-UE interference with proper UE power allocation is beneficial
to optimize the performance objective. Also it should be noted that
$L_{\textrm{algo1}}$ is set to 3 because the proposed algorithm converges
rather quickly with regard to $L_{\textrm{algo1}}$ as have been shown
in Fig.~\ref{fig:Algorithm Convergence}.

\noindent

As future works, closed-form evaluation of the objective function
in the optimization problem, analysis on the relationship between
the performance and feedback size, the impact of CDI feedback delay,
as well as more practical fading channel model and non-RVQ CDI codebook
will be considered for the proposed beamforming schemes. In addition,
the extensions to more sophisticated systems such as MIMO relay networks
and multi-cell cooperative broadcast channels will be investigated
in the future.

\section{Conclusions}

In this paper, leakage-based robust beamforming for MU-MISO system
with per-antenna power constraints and quantized CDI is studied. Based
on our derived CDF of the leakage power for the non-robust ZF beamforming
scheme, we propose the MALC-PA and RALC-PA schemes using a two-stage
algorithm to alternately update the per-UE power allocations and beamforming
vectors in order to maximize the expected weighted sum-rate performance
under per-antenna power constraints. Simulation results show that
the proposed schemes can achieve better performance than the ZF-PA
scheme in terms of average weighted sum-rate performance.

\section*{Appendix I: Proof of Theorem 1}

We decompose $\mathbf{w}_{k}\left(k\neq j\right)$ as $\mathbf{w}_{k}=\sqrt{\tilde{P}_{k}}\mathbf{\tilde{w}}_{k}=\sqrt{\tilde{P}_{k}}\left(\beta_{k,j}\mathbf{\hat{h}}_{j}^{\textrm{H}}+\sqrt{1-\left|\beta_{k,j}\right|^{2}}\mathbf{v}_{k,j}\right)$,
where $\beta_{k,j}=\mathbf{\hat{h}}_{j}\mathbf{\tilde{w}}_{k}$ and
$\mathbf{v}_{k,j}$ is a unit-norm random vector representing the
projection of $\mathbf{\tilde{w}}_{k}$ onto the nullspace of $\mathbf{\hat{h}}_{j}^{\textrm{H}}$.
Then, considering \eqref{eq:decomp_hk_tilde_BS}, we can take the
expectation of $\left|\tilde{\mathbf{h}}_{j}^{\diamond}\mathbf{w}_{k}\right|^{2}$
with respect to $\tilde{\mathbf{h}}_{j}^{\diamond}$ and $\mathbf{\tilde{w}}_{k}$,
and get

\vspace{5pt}

\noindent $\mathbb{E}\left\{ \left|\tilde{\mathbf{h}}_{j}^{\diamond}\mathbf{w}_{k}\right|^{2}\right\} $

\noindent $=\tilde{P}_{k}\mathbb{E}\left\{ \left(1-Z\right)\left|\beta_{k,j}\right|^{2}+Z\left(1-\left|\beta_{k,j}\right|^{2}\right)\left|\mathbf{e}_{j}^{\diamond}\mathbf{v}_{k,j}\right|^{2}\right\} $

\noindent $\;+\tilde{P}_{k}\mathbb{E}\left\{ 2\textrm{Re}\left\{ \sqrt{\left(1-Z\right)Z\left(1-\left|\beta_{k,j}\right|^{2}\right)}\beta_{k,j}\mathbf{e}_{j}^{\diamond}\mathbf{v}_{k,j}\right\} \right\} $

\noindent $\overset{\left(\textrm{a}\right)}{=}\tilde{P}_{k}\mathbb{E}\left\{ \left(1-Z\right)\left|\beta_{k,j}\right|^{2}+Z\left(1-\left|\beta_{k,j}\right|^{2}\right)\left|\mathbf{e}_{j}^{\diamond}\mathbf{v}_{k,j}\right|^{2}\right\} $

\noindent $\overset{\left(\textrm{b}\right)}{=}\tilde{P}_{k}\mathbb{E}\left\{ \left(1-Z\right)\left|\beta_{k,j}\right|^{2}+Z\left(1-\left|\beta_{k,j}\right|^{2}\right)\frac{1}{N-1}\right\} $

\noindent $=\tilde{P}_{k}\left[\frac{\eta}{N-1}+\left(1-\frac{N}{N-1}\eta\right)\mathbb{E}\left\{ \left|\beta_{k,j}\right|^{2}\right\} \right].$

\vspace{-25pt}

\begin{equation}
\qquad\qquad\qquad\qquad\qquad\qquad\quad\qquad\label{eq:E{|hjwk|_squared}}
\end{equation}

\begin{singlespace}
\noindent \begin{flushleft}
\vspace{-20pt}

\par\end{flushleft}
\end{singlespace}

\noindent Equation (a) is obtained because $Z$, $\beta_{k,j}$, $\mathbf{e}_{j}^{\diamond}$
and $\mathbf{v}_{k,j}$ are independently distributed. Equation (b)
holds because $\mathbf{e}_{j}^{\diamond}$ and $\mathbf{v}_{k,j}^{\textrm{H}}$
are i.i.d. isotropic vectors located in the $\left(N-1\right)$-dimensional
nullspace of $\mathbf{\hat{h}}_{j}$. Hence $\left|\mathbf{e}_{j}^{\diamond}\mathbf{v}_{k,j}\right|^{2}$
follows a $\textrm{beta}\left(1,N-2\right)$ distribution, and the
mean value of which equals to $\frac{1}{N-1}$ \cite{integral_table book}.
From \eqref{eq:ineq_posi_def} and \eqref{eq:E{|hjwk|_squared}},
we can see that $\mathbb{E}\left\{ \left|\tilde{\mathbf{h}}_{j}^{\diamond}\mathbf{w}_{k}\right|^{2}\right\} $
is a monotonically increasing affine function with respect to $\mathbb{E}\left\{ \left|\beta_{k,j}\right|^{2}\right\} $.
Thus, $\mathbb{E}\left\{ \left|\tilde{\mathbf{h}}_{j}^{\diamond}\mathbf{w}_{k}\right|^{2}\right\} $
achieves its minimum value

\begin{singlespace}
\noindent
\begin{eqnarray}
\min\left(\mathbb{E}\left\{ \left|\tilde{\mathbf{h}}_{j}^{\diamond}\mathbf{w}_{k}\right|^{2}\right\} \right) & = & \frac{\tilde{P}_{k}\eta}{\left(N-1\right)},\label{eq:min(E{|hjwk|_squared})}
\end{eqnarray}

\end{singlespace}

\noindent when $\mathbb{E}\left\{ \left|\beta_{k,j}\right|^{2}\right\} =0$.
It implies that $\beta_{k,j}=0$ since $\left|\beta_{k,j}\right|^{2}\geq0$
is always true. On the other hand, $\mathbb{E}\left\{ \left|\tilde{\mathbf{h}}_{j}^{\diamond}\mathbf{w}_{k}^{\textrm{ZF}}\right|^{2}\right\} $
can be calculated as

\vspace{5pt}

\noindent $\mathbb{E}\left\{ \left|\tilde{\mathbf{h}}_{j}^{\diamond}\mathbf{w}_{k}^{\textrm{ZF}}\right|^{2}\right\} $\vspace{-5pt}

\begin{singlespace}
\noindent
\begin{eqnarray}
 & = & \mathbb{E}\left\{ \left|\left(\left(\sqrt{1-Z}\mathbf{\hat{h}}_{j}+\sqrt{Z}\mathbf{e}_{j}^{\diamond}\right)\right)\left\Vert \mathbf{w}_{k}^{\textrm{ZF}}\right\Vert \mathbf{\tilde{w}}_{k}^{\textrm{ZF}}\right|^{2}\right\} \nonumber \\
 & = & \tilde{P}_{k}\mathbb{E}\left\{ Z\right\} \mathbb{E}\left\{ \left|\mathbf{e}_{j}^{\diamond}\mathbf{\tilde{w}}_{k}^{\textrm{ZF}}\right|^{2}\right\} \nonumber \\
 & \overset{\left(\textrm{a}\right)}{=} & \frac{\tilde{P}_{k}\eta}{\left(N-1\right)}.\label{eq:E{interf_ZF}}
\end{eqnarray}

\end{singlespace}

\vspace{-5pt}

\noindent Equation (a) comes from the fact that $\left|\mathbf{e}_{j}^{\diamond}\mathbf{\tilde{w}}_{k}^{\textrm{ZF}}\right|^{2}$
also conforms to a $\textrm{beta}\left(1,N-2\right)$ distribution
\cite{Limited-bit CSI}, and the mean value of which is $\frac{1}{N-1}$.
Our proof is completed by combining \eqref{eq:min(E{|hjwk|_squared})}
and \eqref{eq:E{interf_ZF}} to form an inequality with multiplication
of $\xi_{j}^{2}A_{j}$ to both sides since CMI of UE $j$ is independent
of \eqref{eq:E{|hjwk|_squared}}\textasciitilde{}\eqref{eq:E{interf_ZF}}.

\noindent
\begin{figure*}[!tp]
\begin{singlespace}
\noindent
\begin{eqnarray}
P_{D}\left(d\right) & = & \Pr\left(RV\leq d\right)\nonumber \\
 & = & \Pr\left(V\leq\frac{d}{R}\right)\nonumber \\
 & = & \int_{0}^{\infty}P_{V}\left(\frac{d}{r}\right)p_{R}\left(r\right)dr\nonumber \\
 & = & \int_{0}^{d}p_{R}\left(r\right)dr+\int_{d}^{\infty}P_{V}\left(\frac{d}{r}\right)p_{R}\left(r\right)dr\nonumber \\
 & = & 1+\frac{1}{\left(N-2\right)!}\left[\sum_{n=0}^{N-2}\sum_{m=1}^{2^{B}}\textrm{C}_{N-2}^{n}\textrm{C}_{2^{B}}^{m}\frac{\left(-1\right)^{n+m}md^{n}}{mN-\left(m+n\right)}\int_{d}^{\infty}r^{\left(N-1-n\right)}\textrm{e}^{-r}dr\right]\nonumber \\
 &  & -\frac{1}{\left(N-2\right)!}\left[\sum_{n=0}^{N-2}\sum_{m=1}^{2^{B}}\textrm{C}_{N-2}^{n}\textrm{C}_{2^{B}}^{m}\frac{\left(-1\right)^{n+m}md^{m\left(N-1\right)}}{mN-\left(m+n\right)}\int_{d}^{\infty}r^{-\left(m-1\right)\left(N-1\right)}\textrm{e}^{-r}dr\right]\nonumber \\
 & \overset{\left(\textrm{a}\right)}{=} & 1+\frac{1}{\left(N-2\right)!}\left\{ \sum_{n=0}^{N-2}\sum_{m=1}^{2^{B}}\textrm{C}_{N-2}^{n}\textrm{C}_{2^{B}}^{m}\frac{\left(-1\right)^{n+m}md^{n}}{mN-\left(m+n\right)}\left.\left[-\textrm{e}^{-r}\sum_{l=0}^{N-1-n}l!\textrm{C}_{N-1-n}^{l}r^{N-1-n-l}\right]\right|_{r=d}^{r=\infty}\right\} \nonumber \\
 &  & -\frac{1}{\left(N-2\right)!}\left\{ \sum_{n=0}^{N-2}\textrm{C}_{N-2}^{n}\textrm{C}_{2^{B}}^{1}\frac{\left(-1\right)^{n+1}d^{N-1}}{N-1-n}\left.\left[-\textrm{e}^{-r}\right]\right|_{r=d}^{r=\infty}\right\} \nonumber \\
 &  & -\frac{1}{\left(N-2\right)!}\left\{ \sum_{n=0}^{N-2}\sum_{m=2}^{2^{B}}\textrm{C}_{N-2}^{n}\textrm{C}_{2^{B}}^{m}\frac{\left(-1\right)^{n+m}md^{m\left(N-1\right)}}{mN-\left(m+n\right)}\left.\left[\sum_{l=1}^{\left(m-1\right)\left(N-1\right)-1}\frac{\left(-1\right)^{l}\textrm{e}^{-r}}{l!\textrm{C}_{\left(m-1\right)\left(N-1\right)-1}^{l}r^{\left(m-1\right)\left(N-1\right)-l}}\right]\right|_{r=d}^{r=\infty}\right\} \nonumber \\
 &  & -\frac{1}{\left(N-2\right)!}\left\{ \sum_{n=0}^{N-2}\sum_{m=2}^{2^{B}}\textrm{C}_{N-2}^{n}\textrm{C}_{2^{B}}^{m}\frac{\left(-1\right)^{n+m}md^{m\left(N-1\right)}}{mN-\left(m+n\right)}\frac{\left(-1\right)^{\left(m-1\right)\left(N-1\right)-1}}{\left(\left(m-1\right)\left(N-1\right)-1\right)!}\int_{d}^{\infty}\frac{\textrm{e}^{-r}}{r}dr\right\} .\label{eq:proof_theo2}
\end{eqnarray}

\end{singlespace}

\rule[0.5ex]{2.05\columnwidth}{1pt}
\end{figure*}

\section*{Appendix II: Proof of Theorem 2}

According to \eqref{eq:NR_ZF_wk} and \eqref{eq:decomp_hk_tilde_BS},
we can get $\left|\tilde{\mathbf{h}}_{j}^{\diamond}\mathbf{\tilde{w}}_{k}^{\textrm{ZF}}\right|^{2}=\left|\left(\sqrt{1-Z}\mathbf{\hat{h}}_{j}+\sqrt{Z}\mathbf{e}_{j}^{\diamond}\right)\mathbf{\tilde{w}}_{k}^{\textrm{ZF}}\right|^{2}=Z\left|\mathbf{e}_{j}^{\diamond}\mathbf{\tilde{w}}_{k}^{\textrm{ZF}}\right|^{2}.$
Let $G=\left|\mathbf{e}_{j}^{\diamond}\mathbf{\tilde{w}}_{k}^{\textrm{ZF}}\right|^{2}$.
Then we define a random variable $V$ as $V=\left|\tilde{\mathbf{h}}_{j}^{\diamond}\mathbf{\tilde{w}}_{k}^{\textrm{ZF}}\right|^{2}=ZG$.
Furthermore, we define a random variable $D$ as $D=RV$, where the
random variable $R=A_{j}^{\diamond}$. According to \cite{Proakis book},
$R$ is chi-square distributed with $2N$ degrees of freedom, each
with variance $\frac{1}{2}$. Hence its PDF and CDF can be respectively
written as $p_{R}\left(r\right)=\frac{1}{\Gamma\left(N\right)}r^{N-1}\textrm{e}^{-r}$
and $P_{R}\left(r\right)=1-\textrm{e}^{-r}\sum_{l=0}^{N-1}\frac{r^{l}}{l!}$.
Then, based on the result in Theorem~\ref{thm:Theorem_3}, the CDF
of $D=RV$ can be derived as \eqref{eq:proof_theo2}, in which equation
(a) is obtained according to \cite{integral_table book}, where $\int r^{i}\textrm{e}^{-r}dr=-\textrm{e}^{-r}\sum_{l=0}^{i}l!\textrm{C}_{i}^{l}r^{i-l},\left(i\geq0\right)$
and $\int r^{-i}\textrm{e}^{-r}dr=\sum_{l=1}^{i-1}\frac{\left(-1\right)^{l}\textrm{e}^{-r}}{l!\textrm{C}_{i-1}^{l}r^{i-l}}+\frac{\left(-1\right)^{i-1}}{\left(i-1\right)!}\int\frac{\textrm{e}^{-r}}{r}dr,\left(i>0\right).$
With some mathematical manipulations on \eqref{eq:proof_theo2}, we
can get the final form of $P_{D}\left(d\right)$ shown in \eqref{eq:Theorem_2}
of Theorem~\ref{thm:Theorem_2}.

\section*{Appendix III: Proof of Theorem 3}

According to the definitions in Appendix II, $V=\left|\tilde{\mathbf{h}}_{j}^{\diamond}\mathbf{\tilde{w}}_{k}^{\textrm{ZF}}\right|^{2}=ZG$,
where $G=\left|\mathbf{e}_{j}^{\diamond}\mathbf{\tilde{w}}_{k}^{\textrm{ZF}}\right|^{2}$.
According to \cite{Limited-bit CSI}, the probability density function
(PDF) and CDF of $Z$ can be expressed as \vspace{5pt}

\noindent $p_{Z}\left(z\right)=$

\begin{singlespace}
\noindent
\begin{equation}
\left(N-1\right)\sum_{m=1}^{2^{B}}\textrm{C}_{2^{B}}^{m}\left(-1\right)^{m-1}mz^{m\left(N-1\right)-1},\left(z\in\left[0,1\right]\right),
\end{equation}
\vspace{-10pt}

\noindent and\vspace{5pt}

\end{singlespace}

\noindent $P_{Z}\left(z\right)=1-\left(1-z^{N-1}\right)^{2^{B}}$

\begin{singlespace}
\noindent
\begin{equation}
=-\sum_{m=1}^{2^{B}}\textrm{C}_{2^{B}}^{m}\left(-1\right)^{m}z^{m\left(N-1\right)},\left(z\in\left[0,1\right]\right).
\end{equation}

\end{singlespace}

\noindent As explained earlier, $G$ follows a $\textrm{beta}\left(1,N-2\right)$
distribution since $\mathbf{e}_{j}^{\diamond}$ and $\left(\mathbf{\tilde{w}}_{k}^{\textrm{ZF}}\right)^{\textrm{H}}$
are i.i.d. isotropic vectors in the $\left(N-1\right)$-dimensional
nullspace of $\mathbf{\hat{h}}_{j}$ \cite{Limited-bit CSI}. Thus,
the CDF of $G$ is

\noindent $P_{G}\left(g\right)=1-\left(1-g\right)^{N-2}$

\begin{singlespace}
\noindent
\begin{equation}
=-\sum_{n=1}^{N-2}\textrm{C}_{N-2}^{n}\left(-1\right)^{n}g^{n},\left(g\in\left[0,1\right]\right).\qquad\label{eq:PG(g)}
\end{equation}

\end{singlespace}

\noindent Therefore, the CDF of $V=ZG$ can be deduced as \eqref{eq:proof_theo3},
which concludes our proof.

\noindent
\begin{figure*}[!tp]
\noindent
\begin{eqnarray}
P_{V}\left(v\right) & = & \Pr\left(ZG\leq v\right)\nonumber \\
 & = & \Pr\left(G\leq\frac{v}{Z}\right)\nonumber \\
 & = & \int_{0}^{1}P_{G}\left(\frac{v}{z}\right)p_{Z}\left(z\right)dz\nonumber \\
 & = & \int_{0}^{v}p_{Z}\left(z\right)dz+\int_{v}^{1}P_{G}\left(\frac{v}{z}\right)p_{Z}\left(z\right)dz\nonumber \\
 & = & P_{Z}\left(v\right)+\int_{v}^{1}\left[-\sum_{n=1}^{N-2}\textrm{C}_{N-2}^{n}\left(-1\right)^{n}\left(\frac{v}{z}\right)^{n}\right]\left[-\left(N-1\right)\sum_{m=1}^{2^{B}}\textrm{C}_{2^{B}}^{m}\left(-1\right)^{m}mz^{m\left(N-1\right)-1}\right]dz\nonumber \\
 & = & P_{Z}\left(v\right)+\left(N-1\right)\sum_{n=1}^{N-2}\sum_{m=1}^{2^{B}}\textrm{C}_{N-2}^{n}\textrm{C}_{2^{B}}^{m}\left(-1\right)^{n+m}mv^{n}\left.\left[\frac{z^{mN-\left(m+n\right)}}{mN-\left(m+n\right)}\right]\right|_{z=v}^{z=1}\nonumber \\
 & = & 1+\left(N-1\right)\sum_{n=0}^{N-2}\sum_{m=1}^{2^{B}}\textrm{C}_{N-2}^{n}\textrm{C}_{2^{B}}^{m}\left(-1\right)^{n+m}m\frac{v^{n}-v^{m\left(N-1\right)}}{mN-\left(m+n\right)}.\label{eq:proof_theo3}
\end{eqnarray}

\rule[0.5ex]{2.05\columnwidth}{1pt}
\end{figure*}

\end{document}